\overfullrule=0pt
\input harvmac
\def\ulm{{\underline m}}
\def\ulmu{{\underline \mu}}

\def\ulnu{{\underline \nu}}

\def\ulrho{{\underline \rho}}
\def\ulsigma{{\underline \sigma}}
\def\ulhatmu{{\underline {\hat\mu}}}
\def\ulhatnu{{\underline {\hat\nu}}}
\def\ulhatrho{{\underline {\hat\rho}}}
\def\ula{{\underline a}}
\def\ulA{{\underline A}}
\def\ulM{{\underline M}}

\def\ulaa{{\underline \alpha}}
\def\ulb{{\underline b}}
\def\ulbb{{\underline \beta}}
\def\ulc{{\underline c}}
\def\ulgg{{\underline \gamma}}
\def\uld{{\underline d}}
\def\uldd{{\underline \delta}}
\def\a{{\alpha}}

\def\l{{\lambda}}
\def\lt{{\widetilde\lambda}}
\def\b{{\beta}}

\def\g{{\gamma}}
\def\k{{\kappa}}

\def\d{{\delta}}

\def\e{{\epsilon}}

\def\G{{\Gamma}}
\def\L{{\Lambda}}

\def\o{{\omega}}

\def\half{{1\over 2}}
\def\p{{\partial}}

\def\t{{\theta}}

\Title{\vbox{\hbox{IFT-P.002/2002 }}}
{\vbox{
\centerline{\bf Towards Covariant Quantization of the Supermembrane}}}
\bigskip
\centerline{Nathan Berkovits\foot{e-mail: nberkovi@ift.unesp.br}}
\bigskip
\centerline{\it Instituto de F\'\i sica Te\'orica, Universidade Estadual
Paulista}
\centerline{\it Rua Pamplona 145, 01405-900, S\~ao Paulo, SP, Brasil}

\vskip .3in

By replacing ten-dimensional pure spinors with eleven-dimensional
pure spinors, the formalism recently developed for covariantly
quantizing the d=10 superparticle and superstring
is extended to the d=11 superparticle and
supermembrane. In this formalism, kappa symmetry is
replaced by a BRST-like invariance using the nilpotent operator 
$Q=\oint\lambda^\a d_\a$ where $d_\a$ is the worldvolume variable corresponding
to the d=11 spacetime supersymmetric derivative and $\lambda^\a$ is an
SO(10,1) pure spinor variable satisfying $\lambda \Gamma^c \lambda=0$ 
for $c=1$ to 11.

Super-Poincar\'e covariant unintegrated and integrated
supermembrane vertex operators are explicitly constructed which are in
the cohomology of $Q$. After double-dimensional reduction of
the eleventh dimension, these vertex operators are related to Type IIA
superstring vertex operators where $Q=Q_L+Q_R$ is the sum
of the left and right-moving Type IIA BRST operators and
the eleventh component of the pure
spinor constraint, $\lambda \Gamma^{11} \lambda=0$, 
replaces the $b_L^0 -  b_R^0$ constraint of the closed superstring.
A conjecture is made for the computation of M-theory
scattering amplitudes using these supermembrane vertex operators.

\Date {January 2002}

\newsec{Introduction}

There is substantial evidence that d=10 superstring theory contains
nonperturbative symmetries coming from an underlying d=11 theory which
has been named M-theory \ref\hulltown{C.M. Hull and P.K. Townsend,
{\it Unity of String Dualities}, Nucl. Phys. B438 (1995) 109, 
hep-th/9410167\semi
P.K. Townsend, {\it The Eleven-Dimensional Supermembrane Revisited},
Phys. Lett. B350 (1995) 184, hep-th/9501068.}\ref\wittm
{E. Witten, {\it String Theory Dynamics in Various Dimensions},
Nucl. Phys. B443 (1995) 85, hep-th/9503124.}.
It is believed that M-theory properties are 
related to the supermembrane \ref\polch{J. Hughes, J. Liu and
J. Polchinski, {\it Supermembranes}, Phys. Lett. 180B (1986) 370.}
\ref\berg{E. Bergshoeff, E. Sezgin and P. Townsend, {\it
Supermembranes and Eleven-Dimensional Supergravity}, Phys. Lett. B189
(1987) 75.}
\ref\stelle{M. Duff, P. Howe, T. Inami and K. Stelle, {\it Superstrings
in d=10 from Supermembranes in d=11}, Phys. Lett. 191B (1987) 70.}
\ref\duff{M. Duff and J. Lu, {\it Duality Rotations in Membrane
Theory}, Nucl. Phys. B347 (1990) 394.}, 
however, problems with covariantly quantizing
the supermembrane have made it difficult to study these properties. 
Although there exist light-cone methods such as M(atrix) theory\ref\hoppe
{B. deWit, J. Hoppe and H. Nicolai, {\it On the Quantum Mechanics of
Supermembranes}, Nucl. Phys. B305 (1988) 545.}\ref\bss{T. Banks,
W. Fischler, S.H. Shenker and L. Susskind, {\it M Theory as a 
Matrix Model: A Conjecture}, Phys. Rev. D55 (1997) 5112, hep-th/9610043.}
for
studying the supermembrane, the lack of spacetime gauge and Lorentz
symmetries makes these light-cone methods clumsy and limits their use
to special backgrounds. Nevertheless, certain properties of M-theory
have been successfully studied using light-cone supermembrane
vertex operators 
\ref\lcvo{M.B. Green, M. Gutperle and H. Kwon,
{\it Light-cone Quantum Mechanics of the Eleven-dimensional
Superparticle}, JHEP 9908 (1999) 012, hep-th/9907155\semi
A. Dasgupta, H. Nicolai and J. Plefka, {\it
Vertex Operators for the Supermembrane}, JHEP 0005 (2000) 007,
hep-th/0003280\semi
J. Plefka, {\it Vertex Operators for the Supermembrane and Background
Field Matrix Theory}, Int. J. Mod. Phys. A16 (2001) 660, hep-th/0009193\semi
B. Pioline, H. Nicolai, J. Plefka and A. Waldron,
{\it $R^4$ Couplings, the Fundamental Membrane and Exceptional Theta
Correspondence}, JHEP 0103 (2001) 036, hep-th/0102123.},
and it should now be possible to covariantize these light-cone methods
using the results of this paper.

There are two essential problems with covariant quantization of the
supermembrane which, naively, appear to be unrelated. The first problem,
which is also present in the bosonic membrane, is the complicated
non-quadratic
nature of the supermembrane Hamiltonian and the resulting difficulties
in constructing the physical spectrum. The second problem, which is
also present in the Green-Schwarz (GS) superstring \ref\gscov{M.B. Green
and J.H. Schwarz, {\it Covariant Description of Superstrings},
Phys. Lett. B136 (1984) 367.},
is the kappa symmetry \ref\kappas{W. Siegel, {\it
Hidden Local Supersymmetry in the Supersymmetric Particle Action},
Phys. Lett. B128 (1983) 397.} of the supermembrane action which
implies fermionic second-class constraints that are difficult to
covariantly separate out from the first-class constraints.

Recently, a new formalism \ref\pure{N. Berkovits,
{\it Super-Poincar\'e Covariant Quantization of the Superstring},
JHEP 0004 (2000) 018, hep-th/0001035\semi N. Berkovits, {\it Covariant
Quantization of the Superstring}, Int. J. Mod. Phys. A16 (2001) 801,
hep-th/0008145.}\ref\puresuperp{N. Berkovits, {\it Covariant Quantization
of the Superparticle using Pure Spinors}, JHEP 0109 (2001) 016, 
hep-th/0105050.}
was developed for quantizing the superstring
which preserves manifest SO(9,1)
super-Poincar\'e covariance but does not suffer
from the problems of the GS formalism. This formalism uses
a new version of the superstring action which includes 
bosonic pure spinor
ghost variables $\l^\mu$ satisfying $\l\g^m\l=0$ for $m=1$ to 10.\foot
{To simplify d=11 language, the time coordinate will be called $x^{10}$
instead of $x^0$. The indices $m,n,p,...$ and $\mu,\nu,\rho,...$
will label d=10 vectors and spinors, and the indices $a,b,c,...$ and
$\a,\b,\g,...$ will label d=11 vectors and spinors.}
In this pure spinor formalism for the superstring, kappa symmetry is
replaced by a BRST-like invariance using the nilpotent operator 
$Q=\oint \l^\mu d_\mu$ where $d_\mu$ is the worldsheet variable for
the d=10 spacetime supersymmetric derivative. Physical vertex operators are
defined as states in the cohomology of $Q$ and, since the worldsheet 
action is quadratic, manifestly super-Poincar\'e covariant
scattering amplitudes \ref\val{N. Berkovits and B.C. Vallilo,
{\it Consistency of Super-Poincar\'e Covariant Superstring 
Tree Amplitudes}, JHEP 0007 (2000) 015, hep-th/0004171.}
can be computed using the free-field OPE's
of the worldsheet variables.

Since the standard supermembrane action \berg\ reduces to the GS version of the
Type IIA superstring action \gscov\
after double-dimensional reduction of the eleventh
dimension \stelle, it is natural to look for an alternative version of the
supermembrane action which reduces instead to the pure spinor version of the
Type IIA superstring action. Such a generalization 
is reasonable given the results of
\ref\ton{P. Pasti and M. Tonin, {\it Twistor-like Formulation of the
Supermembrane in D=11}, Nucl. Phys. B418 (1994) 337, hep-th/9303156\semi
I.A. Bandos, D. Sorokin, M. Tonin, P. Pasti and D.V. Volkov,
{\it Superstrings and Supermembranes in the Doubly Supersymmetric
Geometrical Approach},
Nucl. Phys. B446 (1995) 79, hep-th/9501113.}
where d=10 twistor-like methods for the superstring
were generalized to d=11 twistor-like methods for the supermembrane.
Furthermore, it was shown by Howe that just as the super-Yang-Mills
equations of motion can be understood as d=10 pure spinor integrability
conditions \ref\howeone{P.S. Howe, {\it Pure Spinor Lines in 
Superspace and Ten-Dimensional Supersymmetric Theories},
Phys. Lett. B258 (1991) 141.}, 
the d=11 supergravity equations of motion can be understood
as d=11 pure spinor integrability conditions \ref\howetwo
{P.S. Howe, {\it Pure Spinors, Function Superspaces and Supergravity
Theories in Ten-Dimensions and Eleven-Dimension}, Phys. Lett. B273 (1991)
90.}.

As will be shown in this paper, it is indeed possible to construct a
supermembrane action which reduces after double-dimensional reduction
to the pure spinor version of the Type IIA superstring action. 
In this pure spinor version of the supermembrane
action, kappa symmetry is replaced by a BRST-like invariance using
the nilpotent operator $Q= \oint\l^\a d_\a$ where 
$d_\a$ is now the worldvolume variable for the d=11 spacetime supersymmetric
derivative and $\l^\a$ is an SO(10,1) pure spinor ghost variable
satisfying $\l\G^c\l=0$ for $c=1$ to 11.\foot{
The definition of d=11 pure spinors used here differs from that of Howe
in \howetwo\ where d=11 pure spinors were required to satisfy
both $\l\G^c\l=0$ and $\l\G^{cd}\l=0$.
Howe's definition of a d=11 pure spinor is more restrictive
than the definition used here and does not appear to be appropriate for
superparticle and supermembrane quantization.} 
After double-dimensional reduction of the
eleventh dimension, $d_\a$ splits into the left and right-moving Type IIA
worldsheet variables $d_{L\mu}$ and $d_{R\hat\mu}$, $\l^\a$ splits into
the left and right-moving Type IIA pure spinor variables $\l_L^\mu$ and
$\l_R^{\hat\mu}$, and $Q$ reduces to the sum of the left and right-moving
Type IIA BRST operators, $Q=Q_L+Q_R$ where $Q_L=\oint \l_L^\mu d_{L\mu}$
and $Q_R=\oint \l_R^{\hat\mu} d_{R\hat\mu}$. 
Furthermore, the eleventh component of the
pure spinor constraint, $\l\G^{11}\l=0$, replaces the $b^0_L-b^0_R$ 
constraint\foot{I 
thank Barton Zwiebach for stressing the importance of the $b^0_L-b^0_R$
constraint.}
which is necessary for defining BRST cohomology in closed string theory
\ref\zwie{B. Zwiebach, {\it Closed String Field Theory: Quantum Action
and the B-V Master Equation}, Nucl. Phys. B390 (1993) 33, hep-th/9206084.}.

Since kappa symmetry is replaced by a BRST-like
invariance, this formalism does not
suffer from quantization problems associated with second-class constraints.
Furthermore, since physical states will be defined as states in the cohomology
of $Q$, the complicated nature of the supermembrane Hamiltonian does not 
directly enter into the computation of the physical spectrum. 
Note that it was proven for the superstring that states in the cohomology
of $Q$ are annihilated by the Hamiltonian \ref\cohom{N. Berkovits,
{\it Cohomology in the Pure Spinor Formalism for the Superstring},
JHEP 0009 (2000) 046, hep-th/0006003\semi N. Berkovits and O. Chand\'{\i}a,  
{\it Lorentz Invariance of the Pure Spinor BRST Cohomology for the 
Superstring}, Phys. Lett. B514 (2001) 394, hep-th/0105149.},
and one expects that this will also be true for the supermembrane.
It might seem surprising that quantization of the supermembrane is
simpler than quantization of the bosonic membrane, but this situation
often occurs in supersymmetric systems 
where second-order differential equations
can be replaced by first-order differential equations.

Before discussing
the supermembrane, it will be useful to first covariantly
quantize
the d=11 superparticle which describes the zero modes of the supermembrane.
To make this paper self-contained, covariant quantization of the
N=1 and N=2 d=10 superparticles 
and Type II superstring will also be reviewed here.

As reviewed in section 2, covariant quantization \puresuperp\ of the N=1 d=10 
superparticle allows a BRST description of super-Maxwell theory 
where the nilpotent BRST operator is $Q=\l^\mu d_\mu$, 
$d_\mu$ is
the N=1 d=10 supersymmetric derivative, and
$\l^\mu$
is a d=10 pure spinor ghost variable satisfying $\l\g^m\l=0$ for $m=1$
to 10.
Using a suitably defined
norm $\langle~~\rangle$ of ghost number three, the super-Maxwell
action can be constructed as $\int d^{10} x \langle \Psi Q\Psi\rangle$
where $\Psi(\l,x,\t)$ is a quantum-mechanical wavefunction depending
on the d=10 pure spinor and superspace variables. At ghost number one,
$\Psi= \l^\mu A_\mu(x,\t)$ where $A_\mu(x,\t)$ describes the super-Maxwell
fields, and at other ghost numbers, $\Psi$ describes the super-Maxwell
ghosts, antifields and antighosts. 

Furthermore, by coupling the pure spinor version of the N=1
d=10 superparticle action to a super-Maxwell background, one obtains the
integrated version of the open superstring massless vertex operator.
By evaluating correlation functions
of these integrated vertex operators 
with the unintegrated massless vertex operators $\Psi=\l^\mu A_\mu$,
one can
compute N=1 d=10 supersymmetric Born-Infeld amplitudes in a manifestly 
super-Poincar\'e covariant manner. The normalization for the worldsheet
zero modes in these correlation functions is defined by the ghost number
three norm used in the super-Maxwell action.

In section 3, the N=2 d=10 superparticle is covariantly quantized using
the BRST operators $Q_L= \l_L^\mu d_{L\mu}$ and
$Q_R= \l_R^{\hat\mu} d_{R\hat\mu}$ where $d_{L\mu}$ and $d_{R\hat\mu}$ are 
the N=2 d=10 supersymmetric derivatives and $\l_L^\mu$ and
$\l_R^{\hat\mu}$ are independent pure spinors satisfying $\l_L\g^m\l_L=
\l_R\g^m\l_R=0$. At non-zero momentum, the physical spectrum corresponds
to linearized Type II supergravity, however, at zero momentum, there are
Type II supergravity states that are missing from the N=2 d=10
superparticle spectrum. This fact is related to the absence of a
$b_L^0-b_R^0$ constraint, which is known from closed string field theory 
\zwie\
to be necessary for obtaining the correct physical spectrum at zero
momentum. The absence of the $b_L^0-b_R^0$ constraint in the N=2
d=10 superparticle also prevents the construction of a
$\int d^{10}x \langle \Psi Q\Psi\rangle$ action for linearized 
Type II supergravity, which is not surprising for the Type IIB
superparticle because of the self-dual five-form field strength
in the spectrum.

In section 4, a pure spinor version of the d=11 superparticle action
is constructed. In this action, kappa symmetry is replaced by a
BRST-like invariance generated by the nilpotent operator $Q=\l^\a d_\a$
where $d_\a$ is the d=11 supersymmetric derivative and $\l^\a$
is a d=11 pure spinor satisfying $\l\G^c\l=0$ for $c=1$ to 11. 
Using the results
of the appendix where the zero momentum cohomology of $Q$ is explicitly
computed, it is argued that the complete cohomology of $Q$ describes
linearized d=11 supergravity \foot{The cohomology of $Q$ was independently
computed in \ref\ceder
{M. Cederwall, B.E.W. Nilsson and D. Tsimpis,
{\it Spinorial Cohomology and Maximally
Supersymmetric Theories}, JHEP 0202 (2002) 009,
hep-th/0110069.}, which appeared a few months
before this preprint was written.
I would like to thank Paul Howe for bringing reference \ceder\ to
my attention.}.
As in the super-Maxwell action constructed
using the N=1 d=10 superparticle, the linearized d=11 supergravity
action can be constructed as $\int d^{11}x \langle \Psi Q\Psi\rangle$
where $\Psi(\l,x,\t)$ is a quantum-mechanical wavefunction depending
on the d=11 pure spinor and superspace variables, and $\langle ~~\rangle$
is a suitably defined norm of ghost number seven. At ghost number three,
$\Psi(\l,x,\t)= \l^\a\l^\b\l^\g A_{\a\b\g}(x,\t)$ where
$A_{\a\b\g}(x,\t)$ describes the linearized d=11 supergravity fields,
and at other ghost numbers, $\Psi$ describes the linearized d=11
supergravity ghosts, antifields and antighosts. The fact that
d=11 supergravity fields carry ghost number three is explained by
their coupling to the three-dimensional worldvolume of the supermembrane,
while the ghost number one super-Maxwell fields and ghost number two
Type II supergravity fields couple respectively to the one-dimensional
superparticle worldline and two-dimensional superstring worldsheet.

When $P_{11}=0$, the d=11 superparticle BRST operator reduces to
$Q=Q_L+Q_R$ where $Q_L$ and $Q_R$ are the N=2 d=10 superparticle 
BRST operators, and the physical spectrum is linearized Type IIA
supergravity without the zero momentum problems that were encountered
using the N=2 d=10 superparticle. This is possible since the 
$b_L^0-b_R^0$ constraint is imposed in the d=11 superparticle by
the eleventh component of the pure spinor constraint, $\l\G^{11}\l=0$,
which is not present in the N=2 d=10 superparticle.

In section 5, covariant quantization of the N=2 d=10 superparticle
is generalized to the Type II superstring by extending the pure
spinor and N=2 d=10 superspace variables to worldsheet fields. After
reviewing the pure spinor version of the closed superstring action in
a flat background, a BRST-invariant action is constructed in a curved
Type II supergravity background where the left and right-moving BRST 
operators, $Q_L=\oint \l_L^\mu d_{L\mu}$ and $Q_R=\oint\l_R^{\hat\mu} 
d_{R\hat\mu}$,
are conserved and nilpotent when the curved background is on-shell 
\ref\howeme{N. Berkovits and P. Howe, {\it Ten-Dimensional Supergravity
Constraints from the Pure Spinor Formalism for the Superstring},
hep-th/0112160.}.

The integrated form of the closed superstring massless
vertex operator is the
linearized contribution of the curved background to the superstring action,
and the unintegrated form of the massless vertex operator is the
N=2 superparticle wavefunction $\Psi(\l_L,\l_R,x,\t_L,\t_R)$. 
Using the left-right product of the zero mode normalization of the
N=1 d=10 superparticle, one can compute Type II superstring massless
tree amplitudes in a manifestly super-Poincar\'e covariant manner
by evaluating the correlation function of these integrated and
unintegrated massless vertex operators on a string worldsheet.
In a flat background, the pure spinor version of the superstring action is
quadratic, so these correlation functions can easily be evaluated using
the free field OPE's of the worldsheet variables.

Finally, in section 6, a pure spinor version of the d=11 supermembrane
action is constructed where kappa symmetry is replaced by
a BRST-like invariance generated by $Q=\oint\l^\a d_\a$ and $\l^\a$ and
$d_\a$ are now worldvolume variables. This supermembrane action reduces
to the pure spinor version of the d=11 superparticle action when the membrane
tension becomes infinite, and reduces to the pure spinor version of the
Type IIA superstring action when the eleventh dimension is compactified
on an infinitesimally small circle keeping the string tension constant.
The supermembrane action is then generalized to a curved d=11 supergravity 
background where the BRST operator $Q=\oint\l^\a d_\a$ is nilpotent
and conserved when the background is on-shell.

As in the superstring, the integrated form of the massless supermembrane
vertex operator is the linearized contribution of the curved background
to the action, and the unintegrated form of the massless supermembrane
vertex operator is the d=11 superparticle wavefunction $\Psi(\l,x,\t)$.
Using the same zero mode normalization as in the d=11 superparticle, 
one can formally define supermembrane scattering amplitudes as correlation
function of these integrated and unintegrated vertex operators on 
a membrane worldvolume. Although the supermembrane action is not quadratic
in a flat background, it might be possible to compute these correlation
functions using a perturbative expansion in the inverse
of the membrane tension. Note
that unlike superstring scattering amplitudes, one does not expect a
genus expansion for supermembrane scattering amplitudes since there
is no coupling of worldvolume curvature to a spacetime field in the
supermembrane action.

In section 7, a conjecture is made
that these supermembrane scattering amplitudes
are M-theory scattering amplitudes which, after compactification of the
eleventh dimension on a circle whose radius depends on the
string coupling constant, reproduce Type IIA superstring scattering amplitudes.
Since the perturbative expansion in the membrane tension preserves
manifest d=11 super-Poincar\'e covariance, these scattering amplitudes
would contain non-perturbative information about the Type IIA superstring
which might be useful for studying M-theory.

\newsec{Covariant Quantization of the N=1 d=10 Superparticle}

Since the d=11 superparticle has a simpler action than the d=11 supermembrane,
it will be useful to explain how to covariantly quantize the d=11
superparticle before discussing the supermembrane. The quantization method
is similar to the method used in \puresuperp\
for quantizing the N=1 d=10 superparticle,
which will be reviewed first. 

\subsec{Standard description of the N=1 d=10 superparticle}

The standard action for the N=1 d=10 superparticle is 
\ref\super{L. Brink and J.H. Schwarz, {\it
Quantum Superspace}, Phys. Lett. 100B (1981) 310.}
\eqn\action{
S=\int d\tau (P_m \Pi^m + e P^m P_m)}
where 
\eqn\defpi{\Pi^m = \dot x^m + {i\over 2} \t^\mu\g^m_{\mu\nu}\dot\t^\nu,}
$m=1$ to 10 is the SO(9,1) vector index with $x^{10}$ as the time
coordinate, $\mu=1$ to 16 is the SO(9,1)
Majorana-Weyl spinor index,
$P_m$ is the canonical momentum for $x^m$, and $e$
is the Lagrange multiplier which enforces the mass-shell condition.
The gamma matrices $\g^m_{\mu\nu}$ and $\g^{m\mu\nu}$ are
$16\times 16$ symmetric matrices which satisfy
$\g^{(m}_{\mu\nu} \g^{n)~ \nu\rho}= 2 \eta^{mn} \d_\mu^\rho$.
Upper spinor indices will denote Weyl d=10 spinors whereas
lower spinor indices will denote anti-Weyl d=10 spinors.
In terms of the standard $32\times 32$ d=10 $\Gamma$-matrices satisfying
$\{\Gamma^m,\Gamma^n\}=2\eta^{mn}$, $\g^m_{\mu\nu}$ and $\g^{m\mu\nu}$
are the off-diagonal blocks of $\Gamma^m$ in the Weyl representation.
Note that any d=10 antisymmetric bispinor $f^{[\mu\nu]}$ can be written
in terms of a three-form as $f^{[\mu\nu]}= (\g_{mnp})^{\mu\nu} f^{mnp}$, and 
any d=10 symmetric bispinor $g^{(\mu\nu)}$ can be decomposed into
a one-form and five-form as 
$g^{(\mu\nu)}=\g_m^{\mu\nu} g^m +(\g_{mnpqr})^{\mu\nu} g^{mnpqr}$.
Furthermore, the d=10 gamma matrices satisfy the identity
$\eta_{mn} \g^m_{(\mu\nu} \g^n_{\rho)\sigma}=0$.

The action of \action\ is invariant under the global N=1 d=10
spacetime-supersymmetry transformations
\eqn\susyone
{\d \t^\mu = \e^\mu, \quad \d 
x^m = {i\over 2}\t\g^m\e, \quad \d P_m = \d e =0,}
and under the local kappa transformations \kappas
\eqn\ka{ \d \t^\mu = P^m (\g_m \k)^\mu,\quad
\d x^m = -{i\over 2}\t\g^m\d\t ,\quad \d P_m =0,\quad \d e= i\dot\t^\nu\k_\nu.}
The canonical momentum to $\t^\mu$, which will be called $p_\mu$, satisfies
$$ p_\mu = \p L/ \p\dot\t^\mu ={i\over 2} P^m (\g_m\t)_\mu,$$
so
canonical
quantization requires that physical states are annihilated by the
sixteen fermionic Dirac constraints defined by
\eqn\dirac{d_\mu = p_\mu -{i\over 2} P_m (\g^m\t)_\mu.}
Since $\{p_\mu,\t^\nu\}=-i\d_\mu^\nu$,
these constraints satisfy the Poisson brackets 
\eqn\antic{\{d_\mu, d_\nu\} = - P_m \g^m_{\mu\nu},}
and since $P^m P_m =0$ is also a
constraint, eight of the sixteen Dirac constraints
are first-class and eight are second-class.
One can easily check that the eight first-class Dirac constraints
generate the kappa transformations of \ka, however, there
is no simple way to covariantly separate out the second-class
constraints. 

Although one cannot covariantly quantize the action of \action, one
can classically couple the superparticle to a super-Maxwell background
using the action
\eqn\actionc{
\widehat S=\int d\tau [P_m \Pi^m + e P^m P_m + q(\dot\t^\mu A_\mu(x,\t)+
\Pi^m A_m(x,\t) ) ]}
where $A_\mu$ and $A_m$ are the spinor and vector super-Maxwell
gauge superfields and $q$ is the charge of the superparticle.
The action of \actionc\ is invariant under spacetime supersymmetry and
under the background gauge transformations $\d A_\mu= D_\mu\L$
and $\d A_m = \p_m\L$ where $D_\mu=
{\p\over{\p\t^\mu}} + {i\over 2}(\g^m\t)_\mu \p_m$.
And if the kappa transformations of \ka\ are modified to
\eqn\kac{ \d \t^\mu = P^m (\g_m \k)^\mu,\quad
\d x^m = -{i\over 2}\t\g^m\d\t ,\quad \d P_m =-q \d\t\g^m W,
\quad \d e= i(\dot\t^\nu +2 ieq W^\nu)\k_\nu}
where $W^\mu = {1\over{10}}\g^{m\mu\nu}(D_\nu A_m-\p_m A_\nu)$
is the super-Maxwell spinor field strength superfield \ref\siegsmax
{W. Siegel, {\it Superfields in Higher Dimensional Space-Time},
Phys. Lett. B80 (1979) 220.}, the
action of \actionc\ is invariant under \kac\ when $A_\mu$ and $A_m$ satisfy
the super-Maxwell equations of motion
$D_\mu A_\nu + D_\nu A_\mu = i\g^m_{\mu\nu} A_m$.

\subsec{Pure spinor description of the N=1 d=10 superparticle}

Instead of using the standard superparticle action of \action, the 
pure spinor formalism for the N=1 d=10 superparticle uses the 
quadratic action \puresuperp
\eqn\pureaction{
S_{pure}=\int d\tau (P_m \dot x^m + p_\mu \dot\t^\mu  +w_\mu\dot\l^\mu 
-\half P^m P_m )}
where $p_\mu$ is now an
independent variable \ref\covmec{W. Siegel, {\it Classical
Superstring Mechanics}, Nucl. Phys. B263 (1986) 93.}, 
$\l^\mu$ is a pure spinor ghost variable satisfying
\eqn\pure{\l\g^m\l=0  \quad {\rm for} ~~m=1~~ {\rm to} ~~10,}
and $w_\mu$ is the canonical momentum to
$\l^\mu$ which is defined up to the gauge transformation 
\eqn\wwgauge{\d w_\mu = (\g^m\l)_\mu \L_m}
for arbitrary gauge parameters $\L_m$. One can easily
show using a $U(5)$ decomposition of a Wick-rotated
SO(10) spinor that the constraint
of \pure\ and the gauge transformation of $w_\mu$ imply that $\l^\mu$
and $w_\mu$ each contain eleven independent components.\foot{Although
the eleven independent components of $\l^\mu$ 
must be complex in order to satisfy \pure, their complex conjugates
$\overline
{\l^\mu}$ never appear in the pure spinor formalism and can therefore be 
ignored.}
The action of \pureaction\ can be written in manifestly spacetime 
supersymmetric notation as 
\eqn\pureactionsusy{
S_{pure}=\int d\tau ( P_m\Pi^m  +d_\mu \dot\t^\mu +w_\mu\dot\l^\mu 
-\half P^m P_m )}
where $\Pi^m $ and $d_\mu$ are defined as in \defpi\ and \dirac. Note
that $d_\mu$ is defined to be invariant under spacetime supersymmetry,
so $p_\mu$ should be defined to transform as $\d p_\mu = {i\over 2}
P_m (\g^m\e)_\mu$ under \susyone. 

To obtain the correct physical spectrum, the action of \pureaction\
needs to be supplemented with the constraint that physical states are in
the cohomology of the BRST-like operator 
\eqn\BRST{Q=\l^\mu d_\mu.}
Note that $Q^2=0$ using \pure\ and \antic,
and carries ghost-number $+1$
if $\l^\mu$ and $w_\mu$ are defined to carry ghost-number $+1$ and $-1$
respectively. Although it is not yet understood how to obtain $Q$
from gauge-fixing a reparameterization invariant action, it is straightforward
to covariantly quantize the superparticle using this BRST operator and check
that one obtains the correct spectrum.

Unlike the usual particle action where the mass-shell condition comes
from the reparameterization constraint $P_m P^m=0$, the mass-shell
condition in the pure spinor formalism is implied indirectly by the
$Q=\l^\mu d_\mu$ constraint. Furthermore, the gauge invariances generated
by $Q$ replace the kappa transformations of \ka\ which are not a symmetry
of \pureaction. Although light-cone gauge fixing is more subtle in
the pure spinor formalism than in the usual formalism, one can check that
the correct counting
of light-cone variables can be obtained by using the pure spinor
ghost variables to cancel the non-physical matter variables. 
The 22 independent bosonic ghost variables of $\l^\mu$ and $w_\mu$ cancel
22 of the 32 fermionic variables of $\t^\mu$ and $p_\mu$, leaving
ten fermionic variables. 
Two of these ten fermionic variables act as the missing $(b,c)$
reparameterization ghosts and cancel the longtitudinal components of
$x^m$ and $P_m$. The remaining eight fermionic variables are the
physical light-cone fermionic variables. 

\subsec{BRST description of super-Maxwell theory}

Using the BRST quantization method, the cohomology of the BRST operator
$Q$ at a fixed ghost number should reproduce the physical fields in
the spectrum. Furthermore, the structure of BRST transformations implies
that if the ghost number of physical fields is defined to be $G$, 
the states at
ghost number less than
$G$ describe spacetime ghosts, the states at ghost number
$G+1$ describe spacetime antifields, and the states at ghost number greater
than $G+1$
describe spacetime antighosts. 
This structure comes from the fact that the BRST transformation of a field
is its gauge transformation using a ghost as the gauge parameter, the BRST
transformation of an antifield is the equation of motion of the corresponding
field, and the BRST transformation of an antighost is the gauge-fixing
condition acting on the antifield.
As will now be reviewed, the cohomology 
of the BRST operator $Q=\l^\mu d_\mu$ for the N=1 d=10 superparticle 
correctly reproduces these states for N=1 d=10 super-Maxwell theory where
the ghost number of physical fields is defined to be $G=1$.

At ghost-number one, the states in the N=1 d=10 superparticle Hilbert
space are described by the wavefunction
\eqn\wavef{\Psi(\l,x,\t) = \l^\mu A_\mu(x,\t)}
where $A_\mu(x,\t)$ is a spinor superfield. Since $\l^\mu\l^\nu$ is
proportional to $(\g_{mnpqr})^{\mu\nu} \l\g^{mnpqr}\l$, 
\eqn\onsh{Q\Psi= \l^\mu\l^\nu D_\nu A_\mu =0}
implies that 
\eqn\eom{(\g_{mnpqr})^{\mu\nu} D_\mu A_\nu=0}
where $D_\mu = 
{\p\over{\p\t^\mu}} + {i\over 2}(\g^m\t)_\mu \p_m$
is the N=1 d=10 supersymmetric derivative
and $mnpqr$ is any five-form direction. And 
\eqn\gau{\d\Psi =Q\L= \l^\mu D_\mu \L}
implies the gauge transformation
\eqn\ginv{\d A_\mu(x,\t) = D_\mu\L(x,\t).}
\eom\ and \ginv\ are the N=1 d=10 super-Maxwell equations of motion and
gauge invariances written
in terms of the spinor gauge superfield $A_\mu(x,\t)$. To see this,
one can expand $A_\mu(x,\t)$ in components as 
\eqn\exp{A_\mu(x,\t) = f_\mu(x) + f_{\mu\nu}(x)\t^\nu + f_{\mu\nu\rho}(x)
\t^\nu\t^\rho + ... .}
Using the gauge invariance and equations of motion of \gau\ and \eom,
one can set 
\eqn\set{f_\mu(x)=0, \quad f_{\mu\nu}(x) = \g^m_{\mu\nu} a_m(x), \quad
f_{\mu\nu\rho}(x) = \eta_{mn}\g^m_{\mu[\nu} \g^n_{\rho]\sigma} \chi^\sigma(x),}
and all higher components of $A_\mu(x,\t)$ to be proportional to 
$a_m(x)$ and $\chi^\rho(x)$, where $a_m(x)$ and $\chi^\rho(x)$ are
the photon and photino satisfying the equations of motion and gauge 
invariances
\eqn\phot{\p^m \p_{[m} a_{n]} = \p^m (\g_m\chi)_\mu =0, \quad
\d a_m = \p_m \omega.}

So the cohomology of $\l^\mu d_\mu$ at ghost number one reproduces
the physical super-Maxwell fields. To check that the cohomology at
other ghost numbers correctly reproduces the super-Maxwell ghosts,
antifields, and antighosts, it is convenient to first compute
the cohomology at zero momentum. As shown in the appendix of \puresuperp,
the cohomology of $Q=\l^\mu d_\mu$ at zero momentum is equivalent to
the cohomology of 
\eqn\defwq{\widetilde Q = \lt^\mu p_\mu+ (\lt\g^m\lt) b_{(-1)m} 
+ c^m_{(1)}
 (\lt\g_m)_\mu u^\mu_{(-1)} + (\lt\g_m\lt)(b^\mu_{(-2)}\g^m_{\mu\nu}
v^\nu_{(1)})}
$$- 2(b_{(-2)\mu} \lt^\mu)(v_{(1)\nu}\lt^\nu)
+ c^\mu_{(2)}(\g_m\lt)_\mu u_{(-2)}^m + (\lt\g^m\lt) v_{(2)m} b_{(-3)}, $$
where $\lt^\mu$ is an unconstrained spinor and $[b_{(-n)},c_{(n)}]$
and $[u_{(-n)},v_{(n)}]$ are new fermionic and bosonic pairs of conjugate
variables of ghost number $[-n,n]$
which cancel the effect of
removing the pure spinor constraint on $\l^\mu$.\foot{This approach
of adding new variables and
removing the pure spinor constraint was recently used to quantize
the superstring in \ref\vannieu
{P.A. Grassi, G. Policastro, M. Porrati and P. Van Nieuwenhuizen,
{\it Covariant Quantization of Superstrings without Pure Spinor
Constraints}, hep-th/0112162.}. 
However, at non-zero momentum, this approach leads
to complications which makes it usefulness unclear. For example,
one needs to include a term $c_{(1)}^m P_m$ in \defwq\ which naively
puts a constraint on the momentum $P_m$.}
The term $b_{(-1)m} (\lt\g^m\lt)$ replaces the pure spinor constraint, and
the other terms in $\widetilde Q$ are needed to eliminate the extra 
gauge invariances implied by this constraint. For example, since
$b_{(-1)m}(\lt\g^m\lt)$ is invariant under $\d b_{(-1)m}=\lt\g^m f$, 
one needs to
include the term $c_{(1)}^m(\lt\g_m)_\mu u^\mu_{(-1)}$. 
And since this term is invariant under
$\d u_{(-1)}^\mu 
= (\lt\g_m\lt)(\g^m f)^\mu - 2\lt^\mu (\lt^\nu f_\nu)$, one
needs to include the term
$(\lt\g_m\lt)(b^\mu_{(-2)}\g^m_{\mu\nu} v^\nu_{(1)}) - 
2(b_{(-2)\mu} \lt^\mu)(v_{(1)\nu}
\lt^\nu)$. 

Since $\lt^\mu$ is unconstrained, it is easy to compute the zero-momentum
cohomology of $\widetilde Q$ at arbitrary ghost number. One finds that
the states in the cohomology are in one-to-one correspondence with the
variables $[1,c_{(1)}^m,v_{(1)\mu},c_{(2)}^\mu, v_{(2)m}, c_{(3)}]$. 
So there is a scalar spacetime ghost
at ghost number zero, a vector and spinor field at ghost number one,
a spinor and vector antifield at ghost number two, and a scalar spacetime  
antighost at ghost number three. This reproduces the desired BRST structure
of super-Maxwell theory since the only gauge field is the photon which implies
a single scalar ghost. Using the map between $Q$ and $\widetilde Q$,
one finds that the corresponding states in the zero-momentum
cohomology of $Q$ with constrained $\l^\mu$ are given by
\eqn\coh{\Psi(\l,\t) = \omega + (\l\g^m\t) a_m + (\l\g^m\t)(\t\g_m \chi)
+ (\l\g^m\t)(\l\g^n\t)(\t\g_{mn}\chi^*) }
$$+ (\l\g^m\t)(\l\g^n\t)(\t\g_{mnp}\t)a^{*p} +
(\l\g^m\t)(\l\g^n\t)(\l\g^p\t)(\t\g_{mnp}\t)\o^*$$
where $\o$ is the spacetime ghost, $a_m$ and $\chi^\mu$ are the fields,
$\chi^*_\mu$ and $a^{*p}$ are the antifields, and $\o^*$ is the antighost.

The cohomology of $Q$ at non-zero momentum can be obtained by finding
the constraints on these component fields implied by $Q\Psi=0$ and
$\d\Psi= Q\L$. One finds that $\o$ and $\o^*$ have trivial cohomology at
non-zero momentum whereas $a^{*p}$ and $\chi^*_\mu$ satisfy the equations
of motion and gauge invariances
\eqn\antieom{\p_p a^{*p}=0, \quad \d a^*_m= \p^n \p_{[m} \sigma_{n]},\quad \d
\chi^*_\mu = \p^q (\g_q\xi)_\mu.}
As expected, the gauge invariances and equations of motion of the 
super-Maxwell antifields
are related to the equations of motion and gauge invariances of the 
super-Maxwell fields
of \phot.

Using the wave function $\Psi$ and the BRST operator $Q=\l^\mu d_\mu$,
one can construct the spacetime action\foot{This action was first
proposed to me by John Schwarz and Edward Witten \ref\jhs{J. Schwarz
and E. Witten, private communication.} and
generalizes to the super-Yang-Mills action
${\cal S}={1\over{g^2}}\int d^{10} x  Tr\langle \Psi Q \Psi + {2\over 3}\Psi^3
\rangle.$ However, there does not appear to be
a non-abelian generalization
for the analogous action constructed using the d=11 superparticle.}
\eqn\spaction{{\cal S}=\int d^{10} x  \langle \Psi Q \Psi \rangle}
where the norm $\langle ~~\rangle$ is defined such that 
\eqn\norm{\langle (\l\g^m\t)(\l\g^n\t)(\l\g^p\t)(\t\g_{mnp}\t) \rangle =1.}
Since
$(\l\g^m\t)(\l\g^n\t)(\l\g^p\t)(\t\g_{mnp}\t)$ is the antighost state in
\coh\ which cannot be written as $Q\L$ for any $\L$, 
the action of \spaction\ is gauge invariant
under $\d\Psi=Q\L$. Furthermore, the equations of motion from varying
$\Psi$ in \spaction\ imply that $Q\Psi=0$ for components in $Q\Psi$ involving
up to five $\t$'s. Although the manifestly supersymmetric equations of
motion require that $Q\Psi=0$ for all components of $Q\Psi$,
one can check that any component of $Q\Psi=0$ with more than five $\t$'s
is an auxiliary equation of motion which does not affect physical fields.
So removing these auxiliary equations of motion changes auxiliary fields
to gauge fields, but does not change the physical content of the theory.

So the action of \spaction\ reproduces the Batalin-Vilkovisky action
for super-Maxwell theory and, if $\Psi$ is restricted to ghost number one,
\spaction\ reproduces the standard super-Maxwell action.
Note that the norm of \norm\ only involves integration over five of
the sixteen $\t^\mu$ variables and therefore resembles a harmonic
superspace. Since there are eleven independent bosonic $\l^\mu$ variables,
one can interpret this integration over five $\t$'s as coming from
a cancellation between the $\l^\mu$ integration and the integration over
eleven of the sixteen $\t^\mu$ variables.

\subsec{Coupling the superparticle to a super-Maxwell background}

To couple the pure spinor version of the superparticle action of
\pureaction\ to a super-Maxwell background in a BRST-invariant manner, it is
convenient to use the Oda-Tonin method of \ref\tonin{I. Oda and
M. Tonin, {\it On the Berkovits Covariant Quantization of GS Superstring},
Phys. Lett. B520 (2001) 398, hep-th/0109051.}
in which one first computes the BRST variation of the standard
superparticle action in a super-Maxwell
background of \actionc. Under the BRST transformation generated by
$Q=\l^\mu d_\mu$, 
\eqn\brsttr{Q\t^\mu = \l^\mu,\quad Q x^m={i\over 2}\l\g^m\t,\quad
Q d_\mu = -i\Pi^m (\g_m\l)_\mu,\quad Qw_\mu= d_\mu,}
where the auxiliary equation of motion $P_m=\Pi_m$ has been used.
One can check that $Q^2$ annihilates all variables except for $w_\mu$,
which satisfies $Q^2 w_\mu=- i\Pi_m(\g^m\l)_\mu$. This is consistent with
the nilpotency of $Q$ since 
$\d w_\mu = -i\Pi_m(\g^m\l)_\mu$ is a gauge transformation of \wwgauge\ with
gauge parameter $\L_m=-i\Pi_m$. 

After fixing the reparameterization gauge $e=-\half$ and using the
auxiliary equation of motion for $P_m$, the standard superparticle
action in a super-Maxwell background of 
\actionc\ transforms under \brsttr\ as
\eqn\acttr{Q\widehat
S=i\int d\tau (\dot\t^\mu -iqW^\mu)(\g^m\l)_\mu \Pi_m}
when the background superfields are on-shell. So if one adds the term
\eqn\adds{S'= \int d\tau ~Q[ (\dot\t^\mu-iqW^\mu) w_\mu]}
to the standard action,
\eqn\resultf{Q(\widehat S+S')=Q\widehat S + \int d\tau~Q^2
[ (\dot\t^\mu-iqW^\mu) w_\mu]
=Q\widehat S + \int d\tau 
(\dot\t^\mu-iqW^\mu) Q^2 w_\mu = 0.}
Therefore, the BRST invariant action for the N=1 d=10 superparticle in
a super-Maxwell background is
\eqn\invac{\widehat S_{pure}= \widehat S+S'}
$$= \int d\tau[\half\Pi^m\Pi_m +d_\mu\dot\t^\mu 
+w_\mu\dot\l^\mu  + q(\dot\t^\mu A_\mu +\Pi^m A_m -id_\mu W^\mu
-i(\l\g^{mn} w) F_{mn})]$$
where $QW^\mu = \l^\nu D_\nu W = \l^\nu (\g^{mn})_\nu{}^\mu F_{mn}$
and $F_{mn}$ is the super-Maxwell superfield whose lowest component is
the vector field strength. 

Furthermore, one can check that the integrand of the super-Maxwell interaction
term,
\eqn\integr{V=
\dot\t^\mu A_\mu +\Pi^m A_m -id_\mu W^\mu
-i(\l\g^{mn} w) F_{mn},}
satisfies $QV = {d\over {d\tau}}(\l^\mu A_\mu)$,
which is the expected relation between the integrated super-Maxwell
vertex operator and the unintegrated ghost number one
vertex operator $\Psi=\l^\mu A_\mu$.
Open superstring massless tree amplitudes can be computed in a manifestly
super-Poincar\'e covariant manner by evaluating the correlation function
of these integrated and unintegrated super-Maxwell vertex operators on
the one-dimensional boundary of an open superstring worldsheet.
By taking the tension of the string to infinity, these amplitudes
reduce to N=1 d=10 supersymmetric Born-Infeld scattering amplitudes.
To compute an N-point tree amplitude,
one needs three unintegrated vertex operators and N-3 integrated
vertex operators, and the normalization for the zero modes is defined
as in \norm\ which, as desired, is non-vanishing for ghost number three.
Furthermore, since the ghost number three antighost state of \norm\ is in
the cohomology of $Q$ and is not the supersymmetric variation of any state
in the cohomology of $Q$, this normalization definition is manifestly
gauge invariant and supersymmetric.

\newsec{Covariant Quantization of the N=2 d=10 Superparticle}

Before quantizing the d=11 superparticle, it will be useful to
discuss the N=2 d=10 superparticle. Since the N=2 d=10 superparticle
describes the zero modes of the Type II superstring, its quantization
is expected to describe a linearized version of Type II supergravity.

As will be seen in this section, there are subtleties at zero momentum 
with quantizing the N=2 d=10 superparticle 
which are related to subtleties
with quantizing the Type II superstring. Recall that there is a left
and right-moving set of $(b_L,c_L)$ and $(b_R,c_R)$ ghosts in
closed string theory, and states in the closed string cohomology
are required to be annihilated by both $Q_L+Q_R$ and by the zero
mode $b^0_L-b^0_R$. The absence of reparameterization ghosts in the pure spinor
formalism makes it difficult to impose the
$b^0_L-b^0_R$ condition, which is related to the difficulty in constructing
a kinetic term for closed superstring field theory. Remarkably,
these subtleties will be resolved for the Type IIA superparticle in
section 4 by taking the $P_{11}=0$ limit of the
d=11 superparticle. The fact that these subtleties are not resolved
for the Type IIB superparticle is not surprising because of the
self-dual five-form field strength in the Type IIB
supergravity spectrum.

\subsec{Standard description of the N=2 d=10 superparticle}

The standard action for the N=2 d=10 superparticle
is
\eqn\actiontwo{
S=\int d\tau (P_m\Pi^m + e P^m P_m)}
where 
\eqn\defpitwo{\Pi^m = \dot x^m + {i\over 2} \t_L^\mu\g^m_{\mu\nu}\dot\t_L^\nu
+ {i\over 2} \t_R^{\hat\mu}\g^m_{\hat\mu\hat\nu}\dot\t_R^{\hat\nu},}
$m=1$ to 10, $\mu=1$ to 16, $\hat\mu=1$ to 16,
and $(\t_L^\mu,\t_R^{\hat\mu})$ are the Type II
fermionic superspace variables. For
the Type IIA superparticle, $\mu$ and $\hat\mu$ denote spinors
of opposite chirality, while for the Type IIB superparticle, 
$\mu$ and $\hat\mu$ denote spinors of the same chirality.

The action of \actiontwo\ is invariant under the global N=2 d=10
spacetime-supersymmetry transformations
\eqn\susytwo
{\d \t_L^\mu = \e_L^\mu, \quad\d\t_R^{\hat\mu}=\e_R^{\hat\mu}, 
\quad \d x^m = {i\over 2}(\t_L\g^m\e_L + \t_R\g^m\e_R), 
\quad \d P_m = \d e =0,}
and under the local kappa transformations
\eqn\katwo{ \d \t_L^\mu = P^m (\g_m \k_L)^\mu,\quad
\d \t_R^{\hat\mu} = P^m (\g_m \k_R)^{\hat\mu},\quad
\d x^m = -{i\over 2}(\t_L\g^m\d\t_L + \t_R\g^m\d\t_R) ,}
$$
\quad \d P_m =0,\quad \d e= i(\dot\t_L^\nu\k_{L\nu}+\dot\t_R^{\hat\nu}
\k_{R\hat\nu}).$$
The canonical momenta to $\t_L^\mu$ and $\t_R^{\hat\mu}$, 
which will be called $p_{L\mu}$ and $p_{R\hat\mu}$, satisfy
$$ p_{L\mu} = \p L/ \p\dot\t_L^\mu ={i\over 2} P^m (\g_m\t_L)_\mu,\quad
p_{R\hat\mu} = \p L/ \p\dot\t_R^{\hat\mu} ={i\over 2} P^m 
(\g_m\t_R)_{\hat\mu},$$
so
canonical
quantization requires that physical states are annihilated by the
32 fermionic Dirac constraints defined by
\eqn\diractwo{d_{L\mu} = p_{L\mu} -{i\over 2} P_m (\g^m\t_L)_\mu, \quad
d_{R\hat\mu} = p_{R\hat\mu} -{i\over 2} P_m (\g^m\t_R)_{\hat\mu}. }
Since $\{p_{L\mu},\t_L^\nu\}= 
-i\d_\mu^\nu$ and
$\{p_{R\hat\mu},\t_R^{\hat\nu}\}= -i\d_{\hat\mu}^{\hat\nu}$,
these constraints satisfy the Poisson brackets 
\eqn\antictwo{\{d_{L\mu}, d_{L\nu}\} = - P_m \g^m_{\mu\nu}, \quad
\{d_{R\hat\mu}, d_{R\hat\nu}\} =  -P_m \g^m_{\hat\mu\hat\nu}, \quad
\{d_{L\mu}, d_{R\hat\nu}\} = 0,}
and since $P^m P_m =0$ is also a
constraint, 16 of the 32 Dirac constraints
are first-class and 16 are second-class.
One can easily check that the 16 first-class Dirac constraints
generate the kappa transformations of \katwo, however, there
is no simple way to covariantly separate out the second-class
constraints. 

\subsec{Pure spinor description of the N=2 d=10 superparticle}

Instead of using the standard N=2 superparticle action of \actiontwo, the 
pure spinor formalism for the N=2 d=10 superparticle uses the 
quadratic action
\eqn\pureactiontwo{
S_{pure}=\int d\tau (P_m \dot x^m + p_{L\mu}\dot\t_L^\mu  +
p_{R\hat\mu}\dot\t_R^{\hat\mu} +w_{L\mu}\dot\l_L^\mu + w_{R\hat\mu}
\dot \l_R^{\hat\mu} -\half P^m P_m )}
where $p_{L\mu}$ and $p_{R\hat\mu}$ are now independent variables, 
$\l_L^\mu$ and $\l_R^{\hat\mu}$ are pure spinor ghost variables satisfying
\eqn\puretwo{\l_L\g^m\l_L=0 \quad {\rm and}\quad \l_R\g^m\l_R=0
 \quad {\rm for} ~~m=1~~ {\rm to} ~~10,}
and $w_{L\mu}$ and $w_{R\hat\mu}$ are defined up to the gauge transformations
\eqn\gaw{\d w_{L\mu} = (\g^m\l_L)_\mu \L_{Lm},\quad
\d w_{R\hat\mu} = (\g^m\l_R)_{\hat\mu} \L_{Rm}, }
for arbitrary gauge parameters $\L_{Lm}$ and $\L_{Rm}$.
The action of \pureactiontwo\ can be written in manifestly spacetime 
supersymmetric notation as 
\eqn\pureactionsusytwo{
S_{pure}=\int d\tau (P_m \Pi^m +d_{L\mu} \dot\t_L^\mu 
+d_{R\hat\mu}\dot\t_R^{\hat\mu}  +w_{L\mu}\dot\l_L^\mu  +w_{R\hat\mu}
\dot\l_R^{\hat\mu} -\half P^m P_m )}
where $\Pi^m $, $d_{L\mu}$ and $d_{R\hat\mu}$
are defined as in \defpitwo\ and \diractwo.

To obtain the correct physical spectrum, the action of \pureactiontwo\
will be supplemented with the constraint that physical states are in
the cohomology of the left and right-moving BRST-like operators
\eqn\BRSTtwo
{Q_L=\l_L^\mu d_{L\mu} \quad {\rm and}\quad Q_R=\l_R^{\hat\mu} d_{R\hat\mu}.}
In other words, physical states $\Psi$ will be defined by the equations
of motion and gauge invariances
\eqn\cohomtwo{Q_L\Psi=Q_R\Psi=0,
\quad \d\Psi= Q_L\L_L + Q_R\L_R,}
where the gauge parameters $\L_L$ and $\L_R$ are constrained to satisfy
$Q_R\L_L = Q_L\L_R=0$. As will now be shown, this definition of physical
states at (left,right) ghost number $(1,1)$ and non-zero momentum
reproduces the correct linearized Type II supergravity spectrum. However,
at zero momentum, the definition of \cohomtwo\ omits certain states in
the supergravity spectrum. As will be explained below, this is caused
by the absence of the N=2 superparticle analog of the $b_L^0-b_R^0$
constraint for the Type II superstring. 

Note that in light-cone gauge, the 44 independent $(\l_L^\mu,\l_R^{\hat\mu})$
and $(w_{L\mu}, w_{R\hat\mu})$ bosonic ghost variables cancel 44 of the 64
fermionic $(\t_L^\mu,\t_R^{\hat\mu})$ and $(p_{L\mu},p_{R\hat\mu})$ variables,
leaving twenty fermionic variables. Two of these twenty fermionic variables
act as the missing $(b,c)$ reparameterization ghosts and cancel the
longtitudinal components of $x^m$ and $P_m$. However, besides the 
sixteen physical light-cone fermionic variables, there are still two
extra fermionic variables which need to be eliminated. These two extra
fermionic variables are the N=2 superparticle analog of the 
$(b_L-b_R, c_L-c_R)$ zero modes in the closed superstring.

\subsec{BRST description of linearized Type II supergravity at 
non-zero momentum}

Since $Q_L$ and $Q_R$ are constructed from independent variables, the
physical states defined by \cohomtwo\ for the N=2 superparticle are
described by the ``left-right'' product of two N=1 superparticle physical
states. At (left,right) ghost number $(1,1)$, the N=2
superparticle wavefunction 
\eqn\twowf{\Psi(\l_L,\l_R,x,\t_L,\t_R)= 
\l_L^\mu \l_R^{\hat\nu} A_{\mu\hat\nu} (x,\t_L,\t_R)}
is physical if $A_{\mu\hat\nu}$ satisfies
\eqn\eomtwo{ (\g_{mnpqr})^{\rho\mu} D_{L\rho} A_{\mu\hat\nu}=
(\g_{mnpqr})^{\hat\rho\hat\nu} D_{R\hat\rho} A_{\mu\hat\nu}=0}
with the gauge invariances
\eqn\gatwo{\d A_{\mu\hat\nu}= D_{L\mu}\L_{L\hat\nu} + D_{R\hat\nu}
\L_{R\mu}\quad
{\rm where}\quad 
(\g_{mnpqr})^{\hat\rho\hat\mu} D_{R\hat\rho} \L_{L\hat\mu} =
(\g_{mnpqr})^{\rho\mu} D_{L\rho} \L_{R\mu} =0,}
and 
\eqn\derivtwo{D_{L\mu}=
{\p\over{\p\t_L^\mu}} + {i\over 2}(\g^m\t_L)_\mu \p_m, \quad
D_{R\hat\mu}=
{\p\over{\p\t_R^{\hat\mu}}} + {i\over 2}(\g^m\t_R)_{\hat\mu} \p_m, }
are the N=2 d=10 supersymmetric derivatives. 

In components, \eomtwo\ and \gatwo\ imply that the physical states
of the N=2 superparticle are described by left-right product of
super-Maxwell photons and photinos. That is, at ghost number $(1,1)$
\eqn\psitwo{\Psi(\l_L,\l_R,x,\t_L,\t_R) = 
(\l_L\g^m\t_L)(\l_R\g^n\t_R) a_{mn}(x)}
$$ +
(\l_L\g^m\t_L)(\l_R\g^n\t_R)(\t_R\g_n)_{\hat\mu} \chi_{L m}^{\hat\mu}(x) +
(\l_L\g^m\t_L)(\t_L\g_m)_\mu(\l_R\g^n\t_R) \chi_{R n}^\mu(x)$$
$$+
(\l_L\g^m\t_L)(\t_L\g_m)_\mu(\l_R\g^n\t_R)(\t_R\g_n)_{\hat\nu}
 F^{\mu\hat\nu}(x) + ...$$
where the higher components in $...$ can be expressed in terms of
$a_{mn}$, $\chi_{Lm}^{\hat\mu}$, $\chi_{Rn}^\mu$ and $F^{\mu\hat\nu}$.
Furthermore, these fields satisfy the equations of motion 
\eqn\compeom{\p^m(\p_m a_{np} -\p_n a_{mp})= \p^m(\p_m a_{np}- 
\p_p a_{nm})=0,}
$$\p^m \p_{[m}\chi_{L n]}^{\hat \mu} = \g^m_{\hat\mu\hat\nu} 
\p_m \chi_{L n}^{\hat\nu}=0,\quad 
\p^m \p_{[m}\chi_{R n]}^\mu = 
 \g^m_{\mu\nu} \p_m \chi_{R n}^\nu=0,$$
$$ \g^m_{\hat\mu\hat\nu} \p_m F^{\rho\hat\nu}
 = \g^m_{\mu\nu} \p_m F^{\nu\hat\rho} =0,$$
and gauge invariances
\eqn\compga{\d a_{mn}= \p_m \omega_{Ln}+\p_n\omega_{Rm},\quad
\d\chi_{Lm}^{\hat\mu}= \p_m\sigma_L^{\hat\mu},\quad
\d\chi_{Rm}^\mu= \p_m\sigma_R^\mu,}
where the gauge parameters satisfy
\eqn\compgasat{
\p^m \p_{[m}\omega_{Ln]}=
\p^m \p_{[m}\omega_{Rn]}= \g^m_{\hat\mu\hat\nu}\p_m\sigma_L^{\hat\nu}=
\g^m_{\mu\nu}\p_m\sigma_R^\nu=0.}

If one chooses the Lorentz gauge 
\eqn\lorentz{\p^m a_{mn}= \p^n a_{mn} = \p^m \chi_{Lm}^{\hat\mu}
= \p^m \chi_{Rm}^\mu
=0,}
the equations of motion and gauge invariances of \compeom\ and \compga\
are those of linearized Type II supergravity where
$a_{mn}= h_{mn}+ b_{mn} + \eta_{mn}\phi$
describes the symmetric
traceless graviton $h_{mn}$,
antisymmetric two-form $b_{mn}$ and dilaton $\phi$, where
$\chi_{Lm}^{\hat\mu}= \rho_{Lm}^{\hat\mu} + \g_m^{\hat\mu\hat\nu}
\xi_{L\hat\nu}$ and
$\chi_{Rm}^\mu= \rho_{Rm}^\mu + \g_m^{\mu\nu}\xi_{R\nu}$
describe
the N=2 gamma-matrix traceless gravitini $[\rho_{Lm}^{\hat\mu},
\rho_{Rm}^\mu]$ and dilatini $[\xi_{L\hat\nu},\xi_{R\nu}]$, 
and where $F^{\mu\hat\nu}$
describes the Ramond-Ramond field strengths.

So at non-zero momentum, where Lorentz gauge
is possible, the ghost number $(1,1)$ fields in $\Psi$ correctly
describe the linearized Type II supergravity fields.
However, at zero momentum, not all the physical Type II supergravity
fields are included in $(a_{mn},\chi_{Lm}^{\hat\mu},\chi_{Rm}^\mu, 
F^{\mu\hat\nu})$.
For example, both the dilaton and the trace of the metric are physical
scalars at zero momentum, but $a_{mn}$ only contains one scalar. 
Similarly, the Ramond-Ramond gauge fields at zero momentum are not
described by $F^{\mu\hat\nu}$. The absence of these zero momentum fields
prevents the construction of a Type II supergravity kinetic term which
would be the N=2 superparticle
analog of the super-Maxwell action constructed 
in \spaction. 

As mentioned earlier, this problem is related to the absence of the
$b_L^0-b_R^0$ constraint in the pure spinor formalism. In closed
string field theory, the correct definition of physical states uses
the BRST cohomology of $Q=Q_L+Q_R$, together with the constraint that
states are annihilated by $b_L^0-b_R^0$. Although this definition agrees
with \cohomtwo\ at non-zero momentum, it does not agree with \cohomtwo\
at zero momentum. Although it will not be possible to impose the
$b_L^0-b_R^0$ constraint for the N=2 d=10 superparticle, it will now
be shown that this constraint is naturally imposed when one quantizes
the d=11 superparticle using pure spinors.

\newsec{Covariant Quantization of the d=11 Superparticle}

In this section, the d=11 superparticle will be covariantly
quantized in a manner which allows a BRST description 
of linearized d=11 supergravity.

\subsec{Standard description of the d=11 superparticle}

The standard action for the d=11 superparticle is 
\eqn\actionel{
S=\int d\tau (P_c\Pi^c  + e P^c P_c)}
where 
\eqn\defpiel{\Pi^c = \dot x^c + {i\over 2} \t^\a\G^c_{\a\b}\dot\t^\b,}
$c=1$ to 11 is the SO(10,1) vector index with $x^{10}$ as the time
coordinate, and $\a=1$ to 32 is the SO(10,1)
spinor index.
The d=11 gamma matrices $\G^c_{\a\b}$ are
$32\times 32$ symmetric matrices which satisfy
$\G^{(c}_{\a\b} \G^{d)~ \b\g}= 2 \eta^{cd} \d_\a^\g$.
In d=11, spinor indices can be raised and lowered using the antisymmetric
metric tensor $C^{\a\b}$ and its inverse $C^{-1}_{\a\b}$.
For example, $\G^{c\a\b}=C^{\a\d}\G^{c\b}_\d= C^{\a\d}C^{\b\g}\G^c_{\d\g}$.
Note that any d=11 antisymmetric bispinor $f^{[\a\b]}$ can be decomposed into
a scalar, three-form, and four-form as
$f^{[\a\b]}= C^{\a\b} f + (\G_{bcd})^{\a\b} f^{bcd} +
(\G_{bcde})^{\a\b} f^{bcde} $,
and 
any d=11 symmetric bispinor $g^{(\a\b)}$ can be decomposed into
a one-form, two-form and five-form as 
$g^{(\a\b)}=\G_c^{\a\b} g^c +
(\G_{cd})^{\a\b} g^{cd} + (\G_{bcdef})^{\a\b} g^{bcdef}$.
Furthermore, the d=11 gamma matrices satisfy the identity
$\eta_{bc} \G^b_{(\a\b} \G^{cd}_{\g\d)}=0$.

The action of \actionel\ is invariant under the global d=11
spacetime-supersymmetry transformations
\eqn\susyel
{\d \t^\a = \e^\a, \quad \d 
x^c = {i\over 2}\t\G^c\e, \quad \d P_c = \d e =0,}
and under the local kappa transformations
\eqn\kael{ \d \t^\a = P^c (\G_c \k)^\a,\quad
\d x^c = -{i\over 2}\t\G^c\d\t ,\quad \d P_c =0,\quad \d e= i\dot\t^\a\k_\a.}
The canonical momentum to $\t^\a$, which will be called $p_\a$, satisfies
$$ p_\a = \p L/ \p\dot\t^\a ={i\over 2} P^c (\G_c\t)_\a,$$
so
canonical
quantization requires that physical states are annihilated by the
32 fermionic Dirac constraints defined by
\eqn\diracel{d_\a = p_\a -{i\over 2} P_c (\G^c\t)_\a.}
Since $\{p_\a,\t^\b\}=-i\d_\a^\b$,
these constraints satisfy the Poisson brackets 
\eqn\anticel{\{d_\a, d_\b\} =-  P_c \G^c_{\a\b},}
and since $P^c P_c =0$ is also a
constraint, 16 of the 32 Dirac constraints
are first-class and 16 are second-class.
One can easily check that the 16 first-class Dirac constraints
generate the kappa transformations of \kael, however, there
is no simple way to covariantly separate out the second-class
constraints. 

\subsec{Pure spinor description of the d=11 superparticle}

At $P_{11}=0$, the action of \actionel\ reduces to the standard Type IIA
N=2 superparticle action of \actiontwo\ where 
$\t_L^\mu = {1\over{\sqrt 2}}(1+\G^{11})_\a^\mu\t^\a$ and
$\t_R^{\hat\mu} 
= {1\over{\sqrt 2}}(1-\G^{11})_\a^{\hat\mu}\t^\a$. This suggests
constructing a new pure spinor version of the d=11 superparticle action
which instead reduces at $P_{11}=0$ to the Type IIA N=2 superparticle action
of \pureactiontwo. This pure spinor version of the d=11 superparticle
action will be defined as the quadratic action
\eqn\pureactionel{
S_{pure}=\int d\tau (P_c\dot x^c +p_\a \dot\t^\a  +w_\a\dot\l^\a 
-\half P^c P_c )}
where $p_\a$ is an independent variable, $\l^\a$ is an SO(10,1)
pure spinor ghost variable satisfying 
\eqn\pureel
{\l\G^c\l=0  \quad {\rm for} ~~c=1~~ {\rm to} ~~11,}
and $w_\a$ is the canonical momentum to
$\l^\a$ which is defined up to the gauge transformation 
\eqn\wwgaugeel{\d w_\a = 
(\G^c\l)_\a \L_c}
for arbitrary gauge parameter $\L_c$. 

With the exception of the d=11 pure spinor constraint of \pureel,
the action of \pureactionel\ reduces when $P_{11}=0$
to the Type IIA N=2 superparticle
action of \pureactiontwo\ where
\eqn\reds{\t_L^\mu = {1\over{\sqrt 2}}(1+\G^{11})_\a^\mu\t^\a,\quad
\t_R^{\hat\mu} = {1\over{\sqrt 2}}(1-\G^{11})_\a^{\hat\mu}\t^\a,}
$$p_{L\mu} = {1\over{\sqrt 2}}(1-\G^{11})^\a_\mu p_\a,\quad
p_{R\hat\mu} = {1\over{\sqrt 2}}(1+\G^{11})^\a_{\hat\mu} p_\a,$$
$$\l_L^\mu = {1\over{\sqrt 2}}(1+\G^{11})_\a^\mu\l^\a,\quad
\l_R^{\hat\mu} = {1\over{\sqrt 2}}(1-\G^{11})_\a^{\hat\mu}\l^\a,$$
$$w_{L\mu} = {1\over{\sqrt 2}}(1-\G^{11})^\a_\mu w_\a,\quad
w_{R\hat\mu} = {1\over{\sqrt 2}}(1+\G^{11})^\a_{\hat\mu} w_\a.$$
However, the d=11 pure spinor constraint $\l\G^c\l=0$ does
not reduce to the N=2 d=10 pure spinor constraints
$\l_L\g^m\l_L=\l_R\g^m\l_R=0$ of \puretwo.
As will now be shown, the difference between
these constraints resolves the difficulties discussed in the previous
subsection for quantization of the Type IIA N=2 superparticle. 

After decomposing $\l^\a$ into $\l_L^\mu$ and $\l_{R\mu}$, the constraint
$\l\G^c\l=0$ for $c=1$ to 11 implies that 
\eqn\pureten{
\l\G^m\l= \l_L\g^m\l_L +\l_R\g^m\l_R=0\quad{\rm for}~~m=1~~{\rm to}~~10,}
$$\l\G^{11}\l = \l_L^\mu \l_{R\mu} = 0.$$
The first line of \pureten\ is obviously satisfied when
$\l_L\g^m\l_L=\l_R\g^m\l_R=0$, however, the second line is new and
is not implied by \puretwo. 

Note that when $\l\G^{11}\l$ is non-zero,
the constraint $\l\G^m\l=0$ implies that 
\eqn\purerem{\l\G^m\G^{11}\l = 
\l_L\g^m\l_L -\l_R\g^m\l_R=0\quad{\rm for}~~m=1~~{\rm to}~~10,}
which are the remaining constraints of \puretwo. 
To prove this, use the d=11 identity $\eta_{de} \G^{cd}_{(\a\b}\G^e_{\g\d)}=0$
to argue that 
\eqn\argue{(\l\G^{c ~11}\l)(\l\G^{11}\l)= -(\l\G^{cm}\l)(\l\G_m\l).}
So when $\l\G^m\l=0$, either $\l\G^{c~11}\l$ or $\l\G^{11}\l$ must vanish.
In the N=2 d=10 superparticle of the previous subsection, $\l\G^{c~11}\l$
was constrained to vanish. However, in the d=11 superparticle, 
$\l\G^{11}\l$ will be constrained to vanish. 

To compute the number of independent degrees of freedom of a
d=11 pure spinor $\l^\a$ satisfying \pureel, note that
$\eta_{mn} \g^m_{(\mu\nu}\g^n_{\kappa)\rho}=0$ implies that 
$h^m = \l_L\g^m\l_L$ satisfies $h^m h_m=0$,
and suppose that $h^m$ is non-vanishing. Then the first line of \pureten\
implies that $\l_R\g^m\l_R=-h^m$, which constrains nine of the
sixteen $\l_{R\mu}$ variables, leaving seven independent variables
for $\l_{R\mu}$. Furthermore, using the argument of the previous
paragraph and assuming that $h^m$ is non-vanishing,
the second line of \pureten\ imposes no further constraints.
So $\l^\a$ contains 23 independent degrees of freedom, sixteen coming
from $\l_L^\mu$ and seven coming from $\l_{R\mu}$.\foot{As in the d=10
case, the d=11 pure spinor $\l^\a$ must be complex to satisfy \pureel,
but its complex conjugate $\overline{\l^\a}$ never appears in the
formalism and can therefore be ignored. Note that if Howe's definition
of d=11 pure spinors of \howetwo\ had been used, $\l^\a$ would have had
sixteen independent components.}

Replacing the N=2 d=10 pure spinor constraint of \puretwo\ with the
d=11 pure spinor constraint of \pureel, one can check that the light-cone
counting of degrees of freedom now gives the correct answer. After using
the 46 independent $\l^\a$ and $w_\a$ bosonic ghost variables to cancel
46 of the 64 fermionic $\t^\a$ and $p_\a$ variables, one is left with
18 fermionic variables. Two of these fermionic variables replace the
missing $(b,c)$ reparameterization ghosts and the remaining 16 
variables are the physical light-cone fermionic variables. So the
$\l\G^{11}\l=0$ condition has effectively replaced the $b_L^0-b_R^0$
constraint of the closed superstring.

Physical states for the d=11 superparticle will be defined as states
in the cohomology of the BRST-like operator 
\eqn\brstel{Q=\l^\a d_\a,}
which is nilpotent because of 
\anticel\ and \pureel. As will now be shown, this definition of 
physical states correctly describes the spacetime fields of linearized
d=11 supergravity, as well as describing the spacetime ghosts, antifields
and antighosts of the theory. Note that $Q$ of \brstel\ reduces at
$P_{11}=0$ to the sum of the N=2 d=10 BRST operators
$Q= Q_R+Q_L= \l_L^\mu d_{L\mu} + \l_{R\mu} d_R^\mu$. However, using the
d=11 superparticle quantization,
the linearized d=11 supergravity states will
reduce at $P_{11}=0$
to the linearized Type IIA supergravity states
without any of the problems at zero momentum
encountered in the previous section using
the N=2 superparticle quantization.

\subsec{BRST description of linearized d=11 supergravity}

As opposed to ghost-number one physical fields of the N=1 d=10
superparticle and ghost-number two physical fields of the N=2 d=10
superparticle, the physical fields of the d=11 superparticle will
appear at ghost-number three in the cohomology of $Q$. 
As will be seen in subsection (7.1), this comes from the fact that 
d=10 super-Maxwell theory couples to the one-dimensional worldline
of the superparticle, Type II supergravity couples to
the
two-dimensional worldsheet of the closed superstring, and d=11
supergravity couples to the three-dimensional worldvolume of the
supermembrane.

At ghost-number three, the d=11 superparticle wavefunction is
\eqn\elewf{\Psi(\l,x,\t)= 
\l^\a \l^\b \l^\g A_{\a\b\g}(x,\t)}
where $A_{\a\b\g}$ is an arbitrary d=11 superfield which is symmetric
in its spinor indices and which, because of \pureel, is defined up to
the algebraic gauge transformation
$\d A_{\a\b\g} = \g^c_{(\a\b} F_{\g) c}$ for arbitrary $F_{\g c}$.
The equation of motion $Q\Psi=0$ implies that  
$\l^\a\l^\b\l^\g\l^\d D_\a A_{\b\g\d}=0$, and $\d\Psi=Q\L$
implies the gauge transformation 
$\d A_{\b\g\d}= D_{(\b}\L_{\g\d)}$ where
$\L=\l^\a\l^\b \L_{\a\b}$ and
$D_\a=
{\p\over{\p\t^\a}} + {i\over 2}(\g^c\t)_\a \p_c$
is the d=11 supersymmetric derivative.
It will now be shown that these equations of motion and gauge invariances
describe the linearized d=11 supergravity fields. In fact, up to
a gauge transformation and with an appropriate choice of conventional
constraints, $A_{\a\b\g}$ is expected\foot
{I would
like to thank Paul Howe for suggesting that $B_{\a\b\g}$ might play
such a role.} to be
the linearized spinor component $B_{\a\b\g}$ of
the three-form superfield $B_{ABC}$ of d=11 supergravity.

Expanding in components, one can show that $\Psi$ of \elewf\ can
be gauge-fixed to the form
\eqn\psiel{\Psi(\l,x,\t) =\l^\a\l^\b\l^\g A_{\a\b\g}(x,\t)}
$$= (\l\g^a\t)(\l\g^b\t)(\l\g^c\t) b_{abc}(x) +
 (\l\g^{(a}\t)(\l\g^{b)c}\t)(\l\g_c\t) g_{ab}(x) $$
$$+ (\l\g^b\t)[(\l\g^c\t)(\l\g^d\t)(\t\g_{cd})_\a -
(\l\g^{cd}\t)(\l\g_c\t)(\t\g_d)_\a] \chi^\a_b(x) + ...$$
where terms in $...$ involve more than four $\t$'s and can
be expressed in terms of $b_{abc}(x)$, $g_{ab}(x)$ and $\chi^\a_b(x)$.
Note that $b_{abc}$ is antisymmetric in its indices and $g_{ab}$
is symmetric in its indices.
Furthermore, the component fields 
$b_{abc}(x)$, $g_{ab}(x)$ and $\chi^\a_b(x)$ satisfy the equations of motion
and gauge invariances\foot{Although one can in principle derive these
equations of motion directly from the higher $\t$ components of $Q\Psi=0$,
they can be justified indirectly using the cohomology structure of
the antifields which will be discussed below.}
\eqn\eqel{\p^d \p_{[a} b_{bcd]}=0,\quad \d b_{abc}= \p_{[a} \omega_{bc]},}
$$\p^a(\p_a g_{bc}-2\p_{(b} g_{c)a}) + \p_b\p_c (\eta^{de}g_{de})=0,\quad
\d g_{ab}= \p_{(a} \rho_{b)},$$
$$(\g^{abc})_{\a\b} \p_b \chi_c^\b=0, \quad \d\chi_a^\b = \p_a\xi^\b,$$
which identify them as the linearized three-form, graviton and gravitino
of d=11 supergravity. So the ghost-number three cohomology of $Q$
correctly describes the linearized d=11 supergravity fields without
any subtleties at zero momentum. 

To show that the cohomology of $Q$ at other ghost numbers correctly
describes the ghosts, antifields and antighosts of linearized d=11
supergravity, it is convenient to first compute the zero momentum
cohomology using a BRST operator $\widetilde Q$
with an unconstrained d=11 spinor
$\widetilde\l^\a$. As discussed in the appendix of this paper, 
the cohomology of $Q$ at
zero momentum is
equivalent to the cohomology of 
\eqn\cozel{\widetilde Q=\lt^\a p_\a +
\lt\G^c\lt b_{(-1)c}+ c_{(1)}^c(\lt\G_{cd}\lt u_{(-2)}^d +
\lt\G^d\lt u_{(-2)[cd]}) + ...}
where $[b_{(-n)},c_{(n)}]$ and 
$[u_{(-n)},v_{(n)}]$ are new fermionic and bosonic pairs of conjugate
variables of ghost number $[-n,n]$ and $...$ involves ghost numbers up
to $[-7,7]$ whose explicit form can be found in the appendix. 
As in the discussion of subsection (2.3) for the N=1 d=10 superparticle,
the term $\lt\G^c\lt b_{(-1)c}$ in \cozel\ imposes the pure spinor
constraint and the terms
$c_{(1)}^c(\lt\G_{cd}\lt u_{(-2)}^d +
\lt\G^d\lt u_{(-2)[cd]}) + ...$ remove the extra gauge invariances implied
by this constraint.

As will be discussed in the appendix, the states in the cohomology of 
$\widetilde Q$ are in one-to-one correspondence with the variables
\eqn\oneto{[1,c_{(1)}^c, v_{(2)}^c, v_{(2)}^{[cd]}, c_{(2)}^\a, c_{(3)}^{(cd)},
c_{(3)}^{[cde]}, v_{(3)}^{c\a}, 
c_{(4)}^{c\a}, v_{(4)}^{[cde]},
v_{(4)}^{(cd)}, v_{(5)}^\a, c_{(5)}^{[cd]}, c^c_{(5)}, v_{(6)}^c, c_{(7)}].}
The ghost number three states corresponding to 
$[c_{(3)}^{(cd)},
c_{(3)}^{[cde]}, v_{(3)}^{c\a}]$ are the graviton, antisymmetric three-form,
and gravitino fields of linearized d=11 supergravity, and the ghost number
four states corresponding to
$[v_{(4)}^{(cd)},
v_{(4)}^{[cde]}, c_{(4)}^{c\a}]$ are their antifields. The ghost number
two states corresponding to 
$[v_{(2)}^c, c_{(2)}^\a, v_{(2)}^{[cd]}]$ are the ghosts coming
from super-reparameterization invariance and the three-form gauge invariance
$\d b_{cde}=\p_{[c}\L_{de]}$, 
and the ghost number five states corresponding to
$[c^c_{(5)}, v_{(5)}^\a, c_{(5)}^{[cd]}]$ are their antighosts.
The ghost number one state corresponding to $c_{(1)}^c$ is the ghost-for-ghost
coming from the gauge invariance of the two-form gauge parameter 
$\d\L_{de}=\p_{[d}\L'_{e]}$, and the ghost number six state corresponding
to $v_{(6)}^c$ is its antighost-for-antighost. Finally, the ghost number
zero state corresponding to $1$ is the ghost-for-ghost-for-ghost
coming from the gauge invariance of the ghost-for-ghost gauge parameter
$\d\L'_e=\p_e\L''$, and the ghost number seven state corresponding to
$c_{(7)}$ is its antighost-for-antighost-for-antighost. So using the
results of the appendix, the zero momentum cohomology of $Q$ correctly
describes the zero momentum ghosts, fields, antifields and antighosts
of linearized d=11 supergravity. 

Using the one-to-one map between states at zero momentum
in the cohomologies of $Q$ and $\widetilde Q$,
one finds that the wavefunction for the zero momentum states in
the cohomology of $Q$ with constrained $\l^\a$ is
\eqn\wfel{\Psi(\l,\t)= \o''+ (\l\t)^c \o'_c + (\l^2 \t^2)^{[cd]}\o_{[cd]}
+(\l^2\t^2)^c\rho_c +(\l^2\t^3)^\a \xi_\a }
$$+(\l^3\t^3)^{[cde]}b_{[cde]} +(\l^3\t^3)^{(cd)} g_{(cd)}+
(\l^3\t^4)^{c\a}\chi_{c\a}$$
$$ +
(\l^4\t^5)^{c\a}\chi^*_{c\a} + 
(\l^4\t^6)^{(cd)} g^*_{(cd)}+
(\l^4\t^6)^{[cde]}b^*_{[cde]} $$
$$+ (\l^5 \t^6)^\a \xi^*_\a + (\l^5\t^7)^c \rho^*_c +
(\l^5\t^7)^{[cd]}\o^*_{[cd]} + (\l^6\t^8)^c \o'^*_c +(\l^7\t^9)\o''^*,$$
where 
$[\o_{[cd]},\o'_c,\o'']$ and
$[\o^*_{[cd]},\o'^*_c,\o''^*]$ are the ghosts and antighosts for
the three-form gauge invariance, 
$[\rho_c,\xi_\a]$ and $[\rho^*_c,\xi^*_\a]$ are the ghosts and
antighosts for the super-reparameterization invariance, 
$[b_{[cde]},g_{(cd)},\chi_{c\a}]$ and
$[b^*_{[cde]},g^*_{(cd)},\chi^*_{c\a}]$ are the linearized
d=11 supergravity fields and antifields, and to simplify notation,
the contractions of the spinor indices of $\l^\a$ and $\t^\b$ in \wfel\
have been suppressed. 
Note that the contractions of the spinor indices in the scalar
ghost number seven state denoted 
$(\l^7\t^9)$ can be determined indirectly using
the fact that
\eqn\elfact{(\l\G^{c_1}\t)(\l\G^{c_2}\t) ... (\l\G^{c_9}\t)=
\e^{c_1 ... c_9 de} (\l\G_{de}\l) (\l^7\t^9),}
which can be proven using the identity $(\l\G^c)_\a (\l\G_{cd}\l)=0$.
This is analogous to the fact that the ghost number three scalar state
$(\l^3\t^5)$ in the N=1 d=10 cohomology satisfies
\eqn\tenfact{(\l\g^{m_1}\t)(\l\g^{m_2}\t) ... (\l\g^{m_5}\t)=
\e^{m_1 ... m_5 npqrs} (\l\G_{npqrs}\l) (\l^3\t^5),}
which can be proven using the identity $(\l\g^m)_\mu (\l\g_{mnpqr}\l)=0$.

The cohomology of $Q$ at non-zero momentum can be obtained by finding
the constraints on the component fields of \wfel\ implied by $Q\Psi=0$ and
$\d\Psi= Q\L$. One finds that all ghosts and antighosts
have trivial cohomology at
non-zero momentum, the supergravity fields satisfy the equations of
motion and gauge invariances of \eqel, and the supergravity antifields
satisfy the equations of motion and gauge invariances
\eqn\elantif{
\p^a b^*_{abc}=0,\quad \d b^*_{abc}=\p^d \p_{[a} \rho_{bcd]},}
$$\p^a g^*_{ab}-\half \p_b (\eta^{de}g^*_{de})=0,\quad
\d g^*_{bc}=\p^a(\p_a \o_{bc}-2\p_{(b} \o_{c)a}) + 
\p_b\p_c (\eta^{de}\o_{de}),$$
$$\p^a\chi^*_{a\b}=0,\quad \d\chi^*_{a\b}=(\g_{abc})_{\a\b} \p^b \xi^{c\a}.$$
As expected, the gauge invariances and equations of motion of the 
d=11 supergravity antifields
are related to the equations of motion and gauge invariances of the 
d=11 supergravity fields
of \eqel.

Using the wave function $\Psi$ and the BRST operator $Q=\l^\a d_\a$,
one can construct the spacetime action
\eqn\spactionel{{\cal S}=\int d^{11} x  \langle \Psi Q \Psi \rangle}
where the norm $\langle ~~\rangle$ is defined such that 
\eqn\normel{\langle (\l^7\t^9) \rangle =1.}
Since
$(\l^7\t^9)$ is the scalar antighost state in
\elfact\ which cannot be written as $Q\L$ for any $\L$, 
the action of \spactionel\ is gauge invariant
under $\d\Psi=Q\L$. Furthermore, the equations of motion from varying
$\Psi$ in \spactionel\ imply that $Q\Psi=0$ for components in $Q\Psi$ involving
up to nine $\t$'s. Although the manifestly supersymmetric equations of
motion require that $Q\Psi=0$ for all components of $Q\Psi$,
one can check that any component of $Q\Psi=0$ with more than nine $\t$'s
is an auxiliary equation of motion which does not affect physical fields.
So as in the super-Maxwell action of \spaction,
removing these auxiliary equations of motion changes auxiliary fields
to gauge fields, but does not change the physical content of the theory.

So the action of \spactionel\ reproduces the Batalin-Vilkovisky action
for linearized d=11 supergravity
theory and, if $\Psi$ is restricted to ghost number three,
\spaction\ reproduces the standard linearized d=11 supergravity action.
Note that the norm of \normel\ only involves integration over nine of
the 32 $\t^\a$ variables and therefore resembles a harmonic
superspace. Since there are 23 independent bosonic $\l^\a$ variables,
one can interpret this integration over nine $\t$'s as coming from
a cancellation between the $\l^\a$ integration and the integration over
23 of the 32 $\t^\a$ variables.

\newsec{Covariant Quantization of the Type II Superstring}

In this section, the pure spinor description of the Type II superstring
will be reviewed using language which will be convenient for generalization
to the supermembrane.

\subsec{Standard description of the Type II superstring}

Using notation similar to that of the N=2 d=10 superparticle action
of \actiontwo,
the standard action for the Type II superstring can be written as
\eqn\iia{S = \int d\tau_0 d\tau_1 (P_m\Pi_0^m  + B^{flat}_{MN}
\p_{[0}Z^M\p_{1]}Z^N
+ e^0 (P^m P_m + \Pi_1^m\Pi_{1 m}) + e^1 P_m \Pi_1^m)}
where 
\eqn\piii{\Pi_0^m = \p_0 x^m +{i\over 2}(\t_L\g^m\p_0\t_L+\t_R\g^m\p_0\t_R),
\quad\Pi_1^m = \p_1 x^m +{i\over 2}(\t_L\g^m\p_1\t_L+\t_R\g^m\p_1\t_R),}
$e^0$ and $e^1$ are Lagrange multipliers for the worldsheet reparameterization 
constraints, $B^{flat}_{MN}$ is the flat value of the Type IIA two-form
superfield, $Z^M=(x^m,\t_L^\mu,\t_R^{\hat\mu})$, and 
$M=(m,\mu,\hat\mu)$ is a d=10
superspace coordinate. Note that after integrating
out $P^m$, $e^0$ and $e^1$,
the action of \iia\ reduces to the usual Nambu-Goto form of
the GS superstring action.

Like the N=2 d=10 superparticle action of \actiontwo, the superstring
action of \iia\ is invariant under global N=2 d=10 supersymmetry 
transformations and under ``left-moving'' and ``right-moving''
kappa transformations. To check kappa symmetry, note that under the
local transformation 
\eqn\iiak{\d\t_L^\mu = \xi_L^\mu,\quad
\d\t_R^{\hat\mu} = \xi_R^{\hat\mu},\quad
\d x^m = -{i\over 2}(\t_L\g^m\xi_L
+\t_R\g^m\xi_R),}
$$ \d P_m = -i( \xi_L\g_m\p_1\t_L-
\xi_R\g_m\p_1\t_R),$$
the action of \iia\ transforms as 
\eqn\triiak{\d S = i \int d^2\tau [(\xi_L\g^m\p_R\t_L)
(P_m -\Pi_{1m})+ 
(\xi_R\g^m\p_L\t_R)(P_m +\Pi_{1m})]}
where $\p_R = \p_0 +(e^1-2e^0) \p_1$
and $\p_L = \p_0 +(e^1+2e^0) \p_1$

So if 
\eqn\ifk{\xi_L^\mu= (P_m-\Pi_{1m})(\g^m\kappa_L)^\mu \quad {\rm and}\quad
\xi_R^{\hat\mu}= (P_m+\Pi_{1m})(\g^m\kappa_R)^{\hat\mu}}
for some $\kappa_{L\mu}$ and $\kappa_{R\hat\mu}$, 
\eqn\changeii{\d S= i\int d^2\tau[(\kappa_{L\nu}
\p_R\t_L^\nu )(P-\Pi_1)^2 +
(\kappa_{R\hat\nu}
\p_L\t_R^{\hat\nu})(P+\Pi_1)^2 ],}
which is cancelled by defining $e^0$ and $e^1$ to transform as
\eqn\latr{\d e^0 = -i
\kappa_{L\nu}
\p_R\t_L^\nu -i \kappa_{R\hat\nu}
\p_L\t_R^{\hat\nu},\quad
\d e^1 = 2i \kappa_{L\nu}
\p_R\t_L^\nu -2i \kappa_{R\hat\nu}\p_L\t_R^{\hat\nu}.}

The canonical momenta to $\t_L^\mu$ and $\t_R^{\hat\mu}$, 
which will be called $p_{L\mu}$ and $p_{R\hat\mu}$, satisfy
$$ p_{L\mu} = \p L/ \p\dot\t_L^\mu ={i\over 2} P^m (\g_m\t_L)_\mu
-B_{\mu N}^{flat}\p_1 Z^N,$$
$$ p_{R\hat\mu} = \p L/ \p\dot\t_R^{\hat\mu} ={i\over 2} P^m 
(\g_m\t_R)_{\hat\mu}
-B_{\hat\mu N}^{flat}\p_1 Z^N ,$$
so
canonical
quantization requires that physical states are annihilated by the
32 fermionic Dirac constraints defined by
\eqn\diractwo{d_{L\mu} = p_{L\mu} -{i\over 2} P_m (\g^m\t_L)_\mu
+B_{\mu N}^{flat}\p_1 Z^N, \quad
d_{R\hat\mu} = p_{R\hat\mu} -{i\over 2} P_m (\g^m\t_R)_{\hat\mu}
+B_{\hat\mu N}^{flat}\p_1 Z^N. }
Using $\{p_{L\mu},\t_L^\nu\}= 
-i\d_\mu^\nu$,
$\{p_{R\hat\mu},\t_R^{\hat\nu}\}= -i\d_{\hat\mu}^{\hat\nu}$,
and the flat space value of 
$H^{flat}_{MNP}=\p_{[M}B^{flat}_{NP]}$, one finds that 
these constraints satisfy the Poisson brackets 
\eqn\antictwo{\{d_{L\mu}, d_{L\nu}\} = - (P_m-\Pi_{1m}) \g^m_{\mu\nu}, \quad
\{d_{R\hat\mu}, d_{R\hat\nu}\} =  -(P_m+\Pi_{1m}) \g^m_{\hat\mu\hat\nu}, \quad
\{d_{L\mu}, d_{R\hat\nu}\} = 0.}
And since $(P-\Pi_1)^2 =0$ and
$(P+\Pi_1)^2 =0$ are also
constraints, 16 of the 32 Dirac constraints
are first-class and 16 are second-class.
One can easily check that the 16 first-class Dirac constraints
generate the kappa transformations of \iiak, however, there
is no simple way to covariantly separate out the second-class
constraints. 

Although one cannot covariantly quantize the action of \iia, one can
classically couple the superstring to a Type II supergravity background
using the action
\eqn\iicurved{\widehat S= \int d^2\tau
[P_\ulm\Pi_0^{\ulm}  + B_{MN}
\p_{[0}Z^M\p_{1]}Z^N
+ e^0 (P^\ulm P_\ulm + \Pi_1^\ulm\Pi_{1 \ulm}) + e^1 P_\ulm \Pi_1^\ulm]}
where $\Pi_0^\ulm = E^\ulm_M\p_0 Z^M$,
$\Pi_1^\ulm = E^\ulm_M\p_1 Z^M$,
[$E^M_\ulm$,$E^M_\ulmu$,$E^M_\ulhatmu$] is the super-vierbein,
[$E_M^\ulm$, $E_M^\ulmu$,$E_M^\ulhatmu$] is the inverse super-vierbein,
$B_{MN}$ is the curved Type II two-form superfield, $M=[m,\mu,\hat\mu]$
denote curved vector and spinor indices, and the underlined
indices $\ulM=[\ulm,\ulmu,\ulhatmu]$ denote tangent-space vector
and spinor indices\foot{To avoid confusion, the indices $a,b,c,...$ and
$\a,\b,\g,...$ will be reserved for d=11 indices.}.

This action is invariant under N=2 d=10 super-reparameterizations of the
background as well as under the two-form gauge transformations
$\d B_{MN}=\p_{[M}\L_{N]}$. Under the local transformation
\eqn\localk{\d Z^M = E^M_\ulmu \xi_L^\ulmu + E^M_\ulhatmu \xi_R^\ulhatmu,\quad
\d P_\ulm = -i(\xi_L\g_\ulm)_\ulmu E^\ulmu_M\p_1 Z^M +i
(\xi_R\g_\ulm)_\ulhatmu E^\ulhatmu_M\p_1 Z^M, }
the action transforms as 
\eqn\triiakc{\d\widehat S = i \int d^2\tau [(\xi_L\g^\ulm)_\ulmu
E^\ulmu_M \p_R Z^M 
(P_\ulm -\Pi_{1\ulm})
+ 
(\xi_R\g^\ulm)_\ulhatmu E^\ulhatmu_M
\p_L Z^M (P_\ulm +\Pi_{1\ulm})]}
when the background superfields are on-shell. So the action is invariant
under kappa symmetry if one defines
\eqn\latrc{\d e^0 = -i
\kappa_{L\ulnu} E^\ulnu_M \p_R Z^M
-i \kappa_{R\ulhatnu} E^\ulhatnu_M \p_L Z^M, \quad
\d e^1 = 2i \kappa_{L\ulnu} E^\ulnu_M
\p_R Z^M -2i
\kappa_{R\ulhatnu} E^\ulhatnu_M \p_L Z^M,}
where
\eqn\ifkc{\xi_L^\ulmu= (P_\ulm-\Pi_{1\ulm})(\g^\ulm\kappa_L)^\ulmu\quad
{\rm and}\quad
\xi_R^{\ulhatmu}= (P_\ulm+\Pi_{1\ulm})(\g^\ulm\kappa_R)^{\ulhatmu}}
for some $\kappa_{L\mu}$ and $\kappa_{R\hat\mu}$. 

\subsec{Pure spinor description of the Type IIA superstring}

Using notation similar to that of the pure spinor version of the
N=2 d=10 superparticle, the pure spinor version of the Type II 
superstring action can be written as 
\eqn\spureii{S_{pure}=\int d\tau_0 d\tau_1 [P_m \p_0 x^m +p_{L\mu}
\p_0\t_L^\mu +p_{R\hat\mu}\p_0\t_R^{\hat\mu} +
w_{L\mu}\p_0\l_L^\mu
+w_{R\hat\mu}\p_0\l_R^{\hat\mu}}
$$-\half (P^m P_m + \p_1 x^m \p_1 x_m) + p_{L\mu}
\p_1\t_L^\mu -p_{R\hat\mu}\p_1\t_R^{\hat\mu} +
w_{L\mu}\p_1\l_L^\mu
-w_{R\hat\mu}\p_1\l_R^{\hat\mu}$$
$$+e_1( P_m \p_1 x^m +p_{L\mu}
\p_1\t_L^\mu +p_{R\hat\mu}\p_1\t_R^{\hat\mu} +
w_{L\mu}\p_1\l_L^\mu
+w_{R\hat\mu}\p_1\l_R^{\hat\mu})]$$
where $p_{L\mu}$ and $p_{R\hat\mu}$ are now independent variables \covmec, 
$\l_L^\mu$ and $\l_R^{\hat\mu}$ are pure spinor ghost variables satisfying
\eqn\puretwo{\l_L\g^m\l_L=0 \quad {\rm and}\quad \l_R\g^m\l_R=0
 \quad {\rm for} ~~m=1~~ {\rm to} ~~10,}
and $w_{L\mu}$ and $w_{R\hat\mu}$ are defined up to the gauge transformations
\eqn\gaw{\d w_{L\mu} = (\g^m\l_L)_\mu \L_{Lm},\quad
\d w_{R\hat\mu} = (\g^m\l_R)_{\hat\mu} \L_{Rm}, }
for arbitrary gauge parameters $\L_{Lm}$ and $\L_{Rm}$.
The action of \spureii\ is quadratic in conformal gauge where $e^1$ is
gauged to zero, but for later comparison with the supermembrane
action, it will be useful to leave $e^1$ in the action and not fix
reparameterization invariance of the $\tau_1$ coordinate. Note, however,
that like the pure spinor version of the superparticle actions, 
reparameterization invariance of the $\tau_0$ coordinate has been fixed in 
\spureii.
The action of \spureii\ can be written in manifestly spacetime 
supersymmetric notation as 
\eqn\spureiisusy{S_{pure}=\int d^2\tau [\widetilde P_m \Pi_0^m +
B_{MN}^{flat} \p_{[0}Z^M\p_{1]}Z^N +d_{L\mu}
\p_0\t_L^\mu +d_{R\hat\mu}\p_0\t_R^{\hat\mu} +
w_{L\mu}\p_0\l_L^\mu
+w_{R\hat\mu}\p_0\l_R^{\hat\mu}}
$$-\half (\widetilde P^m \widetilde P_m + \Pi_1^m \Pi_{1m}) + d_{L\mu}
\p_1\t_L^\mu -d_{R\hat\mu}\p_1\t_R^{\hat\mu} +
w_{L\mu}\p_1\l_L^\mu
-w_{R\hat\mu}\p_1\l_R^{\hat\mu}$$
$$+e_1( \widetilde P_m \Pi_1^m +d_{L\mu}
\p_1\t_L^\mu +d_{R\hat\mu}\p_1\t_R^{\hat\mu} +
w_{L\mu}\p_1\l_L^\mu
+w_{R\hat\mu}\p_1\l_R^{\hat\mu})]$$
where $\Pi^m$, $d_{L\mu}$ and $d_{R\hat\mu}$ are defined as in
\piii\ and \diractwo, $\widetilde P_m = P_m-B^{flat}_{mN}\p_1 Z^N,$ 
and $Z^M=(x^m,\t_L^\mu,\t_R^{\hat\mu})$.
Note that $\widetilde P_m$, $d_{L\mu}$ and $d_{R\hat\mu}$ are 
defined to be invariant under spacetime supersymmetry.

As in the N=2 d=10 superparticle, physical states of the Type II
superstring are defined as states in the cohomology of the 
left and right-moving 
\eqn\BRSTtwoii
{Q_L=\l_L^\mu d_{L\mu} \quad {\rm and}\quad Q_R=\l_R^{\hat\mu} d_{R\hat\mu}.}
In other words, physical states $\Psi$ will be defined by the equations
of motion and gauge invariances
\eqn\cohomtwo{Q_L\Psi=Q_R\Psi=0,
\quad \d\Psi= Q_L\L_L + Q_R\L_R,}
where the gauge parameters $\L_L$ and $\L_R$ are constrained to satisfy
$Q_R\L_L = Q_L\L_R=0$.
Note that $Q_L$ and $Q_R$ are conserved using the equations of motion
\eqn\eomcon{\p_R\l_L^\mu=\p_R d_{L\mu}=0,\quad
\p_L\l_R^{\hat\mu}=\p_L d_{R\hat\mu}=0,}
where $\p_L= \p_0 + (e^1-1)\p_1$ and
$\p_R= \p_0 + (e^1+1)\p_1$. And using \antictwo, one finds that
$Q_L^2=Q_R^2=\{Q_L,Q_R\}=0$.

Since the superstring action of \spureii\
reduces to the N=2 d=10 superparticle
action when all worldsheet variables are independent of $\tau_1$,
the massless sector of the superstring spectrum consists of the Type II
supergravity states found in subsection (3.3). Furthermore, it was shown in
\cohom\ that the massive states in the cohomology of $Q_L$ and $Q_R$ 
reproduce the standard superstring spectrum. 

\subsec{Coupling the superstring to Type II supergravity}

As in subsection (2.4) for the N=1 d=10 superparticle in a super-Maxwell
background, 
the easiest way to obtain the pure spinor version
of the superstring action in a curved Type II supergravity background
is to use the Oda-Tonin method of \tonin\ and first compute the BRST variation
of the standard superstring sigma
model action of \iicurved. Under the BRST transformation
generated by $Q_L=\oint\l_L^\mu d_{L\mu}$ and 
$Q_R=\oint\l_R^{\hat\mu} d_{R\hat\mu}$ in a flat background, 
\eqn\flatiit{Q_L\t_L^\mu=\l_L^\mu,\quad Q_L x^m= {i\over 2}
\l_L\g^m\t_L,\quad Q_L d_{L\mu}=-i\Pi_L^m(\g_m\l_L)_\mu, \quad
Q_L w_{L\mu}=d_{L\mu},}
$$Q_R\t_R^{\hat\mu}=\l_R^{\hat\mu},\quad Q_R x^m= {i\over 2}
\l_R\g^m\t_R,\quad Q_R d_{R\hat\mu}=-i\Pi_R^m(\g_m\l_R)_{\hat\mu}, 
\quad Q_R w_{R\hat\mu}=d_{R\hat\mu},$$
where the equation of motion $\widetilde
P^m= \Pi_0^m+e_1\Pi_1^m$ has been used, 
$\Pi_L^m= \Pi_0^m +(e_1-1)\Pi_1^m$, and 
$\Pi_R^m= \Pi_0^m +(e_1+1)\Pi_1^m$.
In a curved background, these BRST transformations generalize to
\eqn\curviit{\widehat Q_L Z^M=E^M_\ulmu\l_L^\ulmu,\quad \widehat
Q_L d_{L\ulmu}=-i\Pi_L^\ulm(\g_\ulm\l_L)_\ulmu, \quad
\widehat Q_L w_{L\ulmu}=d_{L\ulmu},}
$$\widehat Q_R Z^M=E^M_\ulhatmu\l_R^\ulhatmu,\quad \widehat
Q_R d_{R\ulhatmu}=-i\Pi_R^\ulm(\g_\ulm\l_R)_\ulhatmu, \quad
\widehat Q_R w_{R\ulhatmu}=d_{R\ulhatmu},$$
where $E^M_\ulmu$ and $E^M_\ulhatmu$ are defined as in \iicurved,
$\Pi_L^\ulm= E_M^\ulm \p_L Z^M$ and
$\Pi_R^\ulm= E_M^\ulm \p_R Z^M$.

After fixing the reparameterization gauge $e^0=-\half$ and using
the equation of motion for $P_\ulm$, the standard superstring sigma
model action of
\iicurved\ transforms under \curviit\ as
\eqn\brstiit{\widehat Q_L \widehat S = i\int d^2\tau
(\l_L\g^\ulm)_\ulmu
E^\ulmu_M \p_R Z^M \Pi_{L\ulm}, \quad
\widehat Q_R \widehat S = i\int d^2\tau
(\l_R\g^\ulm)_\ulhatmu E^\ulhatmu_M
\p_L Z^M
\Pi_{R\ulm}}
when the background superfields are on-shell. 
If $\l_L^\ulmu$ and $\l_R^\ulhatmu$ were replaced by
$\xi_L^\ulmu$ and $\xi_R^\ulhatmu$ of \ifkc, this would be a left and
right-moving kappa transformation, which could be cancelled by
shifting $e^0$ and $e^1$ as in \latrc. However, in the pure spinor
formalism, the transformation of \brstiit\ will be cancelled by adding to
the action the term
\eqn\newterm{S' = \int d^2\tau[
\widehat Q_L(w_{L\ulmu} E^\ulmu_M \p_R Z^M )+
\widehat Q_R(w_{R\ulhatmu} E^\ulhatmu_M \p_L Z^M ) -
\widehat Q_L\widehat Q_R(w_{L\ulmu}w_{R\ulhatnu} R^{\ulmu\ulhatnu})]}
where $R^{\ulmu\ulhatnu}$ is a superfield whose lowest component is
$e^\phi$ times the Ramond-Ramond field strength $F^{\ulmu\ulhatnu}$.
Using Bianchi identities and equations of motion, 
one can show that $R^{\ulmu\ulhatnu}$
is related to the superspace torsion 
$T_{\underline M\underline N}^{\underline P}$
by \howeme
\eqn\torsionp{
T_{\ulrho \ulm}^\ulhatnu= i\g_{\ulm\ulrho\ulmu}R^{\ulmu\ulhatnu},
\quad
T_{\ulhatrho \ulm}^\ulmu= -i\g_{\ulm\ulhatrho\ulhatnu}R^{\ulmu\ulhatnu}.}

To see that $\widehat Q_L(\widehat S+S')=0$, note that since
$\widehat Q_L^2$ annihilates all variables except for $w_{L\ulmu}$,
\eqn\checkprime{\widehat Q_L S' = \int d^2\tau[(\widehat Q_L^2 
w_{L\ulmu}) E^\ulmu_M\p_R Z^M - \widehat Q_R (w_{R\ulhatmu}
\widehat Q_L(E^\ulhatmu_M \p_L Z^M))-
\widehat Q_R ((\widehat Q_L^2 w_{L\ulmu}) w_{R\ulhatnu}R^{\ulmu\ulhatnu})]}
$$= 
 \int d^2\tau[ -i\Pi_L^\ulm (\l_L\g_\ulm)_\ulmu E^\ulmu_M \p_R Z^M
-\widehat Q_R (w_{R\ulhatmu} \l_L^\ulnu T_{\ulnu\ulm}^\ulhatmu \Pi_L^\ulm)
-\widehat Q_R(-i\Pi_L^\ulm (\l_L\g_\ulm)_\ulmu w_{R\ulhatnu} 
R^{\ulmu\ulhatnu})]$$
$$=\int d^2\tau[ -i\Pi_L^\ulm (\l_L\g_\ulm)_\ulmu E^\ulmu_M \p_R Z^M]
=-\widehat Q_L \widehat S$$
where the Type II on-shell torsion constraints have been used.
Similarly, one can show that $\widehat Q_R(\widehat S+S')=0$.

So the classically BRST-invariant superstring action in a curved
Type II background is given by $\widehat S_{pure}=
\widehat S+S'$, however, to preserve
quantum BRST invariance \howeme, 
one also needs to add the Fradkin-Tseytlin term
$\a'\int d^2\tau \Phi r$ to the superstring action where $r$ is the
worldsheet curvature and $\Phi(x,\t_L,\t_R)$ is a scalar superfield
whose lowest component is the spacetime dilaton. Using the BRST
transformation of \curviit\
to explicitly compute $S'$ of \newterm, one finds that the
pure spinor action in a curved background is
\eqn\iicpure{\widehat S_{pure}= \int d^2\tau[ \half \Pi_L^\ulm\Pi_{R\ulm}
+ B_{MN}\p_{[0}Z^M\p_{1]}Z^N 
+d_{L\ulmu} E_M^\ulmu \p_R Z^M + d_{R\ulhatmu}
E_M^\ulhatmu \p_L Z^M }
$$ + w_{L\ulmu} \p_R\l_L^\ulmu +
w_{R\ulhatmu}\p_L\l_R^\ulhatmu +\Omega_{M\ulmu}^\ulnu \p_R Z^M
\l_L^\ulmu w_{L\ulnu} +
\widehat\Omega_{M\ulhatmu}^\ulhatnu
\p_L Z^M \l_R^\ulhatmu w_{R\ulhatnu} $$
$$+ R^{\ulmu\ulhatnu} d_{L\ulmu} d_{R\ulhatnu} +C^{\ulhatmu\ulnu}_\ulrho
d_{R\ulhatmu}\l_L^\ulrho w_{L\ulnu}+
\widehat C^{\ulmu\ulhatnu}_\ulhatrho
d_{L\ulmu}\l_R^\ulhatrho w_{R\ulhatnu}+
S^{\ulmu\ulhatnu}_{\ulsigma\ulhatrho} \l_L^\ulsigma w_{L\ulmu}
\l_R^\ulhatrho w_{R\ulhatnu}
+\a' r\Phi]$$
where the explicit relations between the background superfields
appearing in \iicpure\ are explained in \howeme.

By computing the linearized contribution of the background superfields
to $\widehat S_{pure}$ of \iicpure, one obtains the integrated form of the
massless Type II superstring vertex operator $\int d^2\tau V$.
Since $\widehat S_{pure}$ is BRST invariant, $Q_L\int d^2\tau V$ and
$Q_R\int d^2\tau V$ must vanish up to worldsheet equations of motion
when $Q_L$ and $Q_R$ generate the BRST transformations of
\flatiit\ in a flat background. 

Once one has the integrated BRST-invariant
vertex operator associated with a physical state, there is a simple
method for obtaining the unintegrated BRST-invariant vertex operator
associated with this state. Since $(Q_L+Q_R)\int d^2\tau V=0$,
$(Q_L+Q_R)V=\p_i W^i$ for some ghost number one state $W^i$
where $i=0$ or 1. And since $(Q_L+Q_R)^2 V=0$, 
$(Q_L+Q_R)W^i = \epsilon^{ij}\p_j U$ for some ghost number two state
$U$ satisfying $(Q_L+Q_R)U=0$. This ghost number two state $U$ is defined
to be the unintegrated closed superstring vertex operator associated
with the physical state represented by $V$. Using this method, one
finds that the unintegrated massless vertex operator $U$ associated
with the linearized contribution to $\widehat S_{pure}$ is the
ghost number two N=2 d=10 superparticle wavefunction 
$\Psi(\l_L,\l_R,x,\t_L,\t_R)$=$\l_L^\mu\l_R^{\hat\nu}
A_{\mu\hat\nu}(x,\t_L,\t_R)$ of \twowf.

Closed superstring massless $N$-point tree amplitudes can be computed
in a manifestly super-Poincar\'e covariant manner
by evaluating correlation functions of $N-3$ massless integrated 
vertex operators $V$ with three massless unintegrated vertex operators
$U$. The normalization for the worldsheet zero modes is the ``left-right''
product of the norm of \norm, $\langle (\l_L^3\t_L^5)
(\l_R^3\t_R^5)\rangle=1$, which implies that the amplitudes are gauge
invariant and supersymmetric when the external states are on-shell.

\newsec{Covariant Quantization of the Supermembrane}

In this section, the methods developed in the previous sections
for the d=11 superparticle and Type II superstring are generalized to
construct a BRST-invariant action for the supermembrane. 
Almost all of the intuition needed for constructing this action comes
from the requirements that the action reduces to the d=11 superparticle
and Type IIA superstring actions in the appropriate limits. That is,
in the limit where all worldvolume variables are independent of
coordinates $\tau_1$ and $\tau_2$, the supermembrane action must reduce
to the d=11 superparticle action of section 4. And in the limit
when $P_{11}=0$, $x^{11}=\tau_2$, and all other worldvolume variables
are independent of $\tau_2$, the action must reduce to the Type IIA
superstring action of section 5.

\subsec{Standard description of the supermembrane}

Using notation similar to that of the d=11 superparticle action
of \actionel,
the standard action for the supermembrane can be written as
\eqn\sma{S = \int d\tau_0 d\tau_1 d\tau_2 (P_c\Pi_0^c  + B^{flat}_{ABC}
\p_{[0}Z^A\p_{1}Z^B \p_{2]} Z^C
+ e^0 (P^c P_c + \det(\Pi_I^c\Pi_{Jc})) + e^I P_c \Pi_I^c)}
where $I,J=1$ to 2, $\det(\Pi_I^c\Pi_{Jc})= (\Pi_1^c\Pi_{1c})(\Pi_2^d\Pi_{2d})
-(\Pi_1^c \Pi_{2c})^2$, 
$\Pi_0^c = \p_0 x^c +{i\over 2}(\t\G^c\p_0\t)$,
$\Pi_I^c = \p_I x^c +{i\over 2}(\t\G^c\p_I\t)$,
$e^0$ and $e^I$ are Lagrange multipliers for the worldsheet reparameterization 
constraints, $B^{flat}_{ABC}$ is the flat value of the d=11 three-form
superfield, $Z^A=(x^a,\t^\a)$, and 
$A=(a,\a)$ is a d=11
superspace coordinate. Note that after integrating
out $P^m$, $e^0$ and $e^I$,
the action of \sma\ reduces to the usual Nambu-Goto form of
the supermembrane action \berg.

Like the d=11 superparticle action of \actionel, the supermembrane
action of \sma\ is invariant under global d=11 supersymmetry 
transformations and under 
kappa transformations. To check kappa symmetry, note that under the
local transformation 
\eqn\smak{\d\t^\a = \xi^\a,\quad
\d x^c = -{i\over 2}(\t\G^c\xi),\quad
\d P_c = -i( \xi\G_{cd}\p_I\t)\Pi_J^d \e^{IJ},\quad
\d e^I = 2ie^0\e^{IJ} \xi_\a\p_J\t^\a,}
the action of \sma\ transforms as 
\eqn\trsmak{\d S = i \int d^3\tau \xi^\a(\G^c_{\a\b}P_c-\half
\G^{cd}_{\a\b}\Pi_{Ic}\Pi_{Jd}\e^{IJ})\nabla\t^\b}
where 
\eqn\defnab{\nabla\t^\b
= (\p_0 +e^I\p_I)\t^\b-2e^0\G_{e\g}^{\b}\p_I\t^\g\Pi_J^e\e^{IJ}.}

So if 
\eqn\ifsmak{\xi^\a= (\G_c^{\a\b}P^c +\half\G_{cd}^{\a\b}\Pi_I^c\Pi_J^d
\e^{IJ}) \kappa_\b}
for some $\kappa_\b$, 
\eqn\changesma{\d S= i\int d^3\tau[\kappa_\b \nabla\t^\b (P^d P_d+
\det(\Pi_I^c\Pi_{Jc})) + 2 (\kappa\G^c\nabla\t)\Pi_{Ic}
(P^d \Pi_{Jd})\e^{IJ}].}
So $\d S$ can be cancelled if $e^0$ and $e^I$ are defined 
to transform as
\eqn\lsmatr{\d e^0 = -i
\kappa_\b \nabla\t^\b,\quad
\d e^I = 2ie^0\e^{IJ}\xi_\a\p_J\t^\a +2i\e^{IJ}(\kappa\G^c\nabla\t)\Pi_{Jc}}
where the first term in $\d e^I$ comes from the transformation of \smak.

The canonical momenta to $\t^\a$, 
which will be called $p_\a$, satisfies
$$ p_\a = \p L/ \p\dot\t^\a ={i\over 2} P^c (\G_c\t)_\a
-\half B_{\a BC}^{flat}\p_I Z^B \p_J Z^C \e^{IJ},$$
so
canonical
quantization requires that physical states are annihilated by the
32 fermionic Dirac constraints defined by
\eqn\diracsm{d_\a = p_\a -{i\over 2} P_c (\G^c\t)_\a
+\half B_{\a BC}^{flat}\p_I Z^B \p_J Z^C \e^{IJ}.}
Using $\{p_\a,\t^\b\}= 
-i\d_\a^\b$
and the flat space value of 
$H^{flat}_{ABCD}=\p_{[A}B^{flat}_{BCD]}$, one finds that 
these constraints satisfy the Poisson brackets 
\eqn\antsmtwo{\{d_\a, d_\b\} = - P_c \G^c_{\a\b} +\half\e^{IJ}\Pi_{Ic}\Pi_{Jd}
\G^{cd}_{\a\b}.}
And since
\eqn\sinces{(-P_c \G^c_{\a\b} +\half\e^{IJ}\Pi_{Ic}\Pi_{Jd}\G^{cd}_{\a\b})
(P^a \G_a^{\b\g} +\half\e^{KL}\Pi_K^a\Pi_L^b\G_{ab}^{\b\g})}
$$=-\d_\a^\g (P_c P^c +\det(\Pi_I^c\Pi_{Jc})) -2 \e^{IJ}(P^c\Pi_{Ic}) \Pi_{Jd}
\G^{d\g}_\a$$
is proportional to the reparameterization constraints,
16 of the 32 Dirac constraints
are first-class and 16 are second-class.
One can easily check that the 16 first-class Dirac constraints
generate the kappa transformations of \smak, however, there
is no simple way to covariantly separate out the second-class
constraints. 

Although one cannot covariantly quantize the action of \sma, one can
classically couple the supermembrane to a d=11 supergravity background
using the action
\eqn\smacurved{\widehat S= \int d^3\tau
[P_\ulc\Pi_0^{\ulc}  + B_{ABC}
\p_{[0}Z^A\p_{1}Z^B \p_{2]} Z^C
+ e^0 (P^\ulc P_\ulc + \det(\Pi_I^\ulc\Pi_{J\ulc})) + e^I P_\ulc \Pi_I^\ulc]}
where $\Pi_0^\ulc = E^\ulc_A\p_0 Z^A$,
$\Pi_I^\ulc = E^\ulc_A\p_I Z^A$,
$[E^A_\ulc,E^A_\ulaa]$ is the super-vierbein,
$[E_A^\ulc,E_A^\ulaa]$ is the inverse super-vierbein,
$B_{ABC}$ is the curved three-form superfield, $A=[a,\a]$ denote
curved vector and spinor indices, and the underlined
indices $\ulA=[\ulc,\ulaa]$ denote d=11 tangent-space vector
and spinor indices.

This action is invariant under d=11 super-reparameterizations of the
background as well as under the three-form gauge transformations
$\d B_{ABC}=\p_{[A}\L_{BC]}$. Under the local transformation
\eqn\localsmk{\d Z^A = E^A_\ulaa \xi^\ulaa,\quad
\d P_\ulc = -i(\xi\G_{\ulc\uld})_\ulaa E^\ulaa_A\p_I Z^A \Pi_J^\uld\e^{IJ},
\quad \d e^I= 2i e^0\e^{IJ}\xi_\ulaa E^\ulaa_A\p_J Z^A,}
the action transforms as 
\eqn\trsmakc{\d \widehat S = i \int d^3\tau \xi^\ulaa(\G^\ulc_{\ulaa\ulbb}
P_\ulc-\half
\G^{\ulc\uld}_{\ulaa\ulbb}\Pi_{I\ulc}\Pi_{J\uld}\e^{IJ})\nabla \Theta^\ulbb}
when the background is on-shell where 
\eqn\defEab{\nabla \Theta^\ulbb
= E^\ulbb_A (\p_0 +e^I\p_I) Z^A-2e^0\G_{\ula\ulgg}^{\ulbb}E^{\ulgg}_A
\p_I Z^A\Pi_J^\ula\e^{IJ}.}
So the action is invariant
under kappa symmetry if one defines
\eqn\lsmatr{\d e^0 = -i
\kappa_\ulbb \nabla \Theta^\ulbb,\quad
\d e^I = 2ie^0\e^{IJ}
\xi_\ulaa E^\ulaa_A\p_J Z^A +2i\e^{IJ}(\kappa\G^\ulc
\nabla \Theta) \Pi_{J\ulc}}
where
\eqn\ifsmakc{\xi^\ulaa= (\G_\ulc^{\ulaa\ulbb}P^\ulc +
\half\G_{\ulc\uld}^{\ulaa\ulbb}\Pi_I^\ulc\Pi_J^\uld
\e^{IJ}) \kappa_\ulbb}
for some $\kappa_\ulbb$. 

\subsec{Pure spinor description of the supermembrane}

Using worldvolume variables which generalize the worldline variables
of the pure spinor version of the d=11 superparticle, 
the pure spinor version of the supermembrane action will be defined as
\eqn\spuresma{S_{pure}=\int d\tau_0 d\tau_1 d\tau_2 [\widetilde
P_c \Pi^c_0 + B^{flat}_{ABC} \p_{[0}Z^A \p_1 Z^B \p_{2]} Z^C  
+d_\a \p_0\t^\a + w_\a\p_0\l^\a}
$$-\half (\widetilde P^c \widetilde P_c + \det(\Pi_I^c\Pi_{Jc})) + 
(d \G_c\p_I\t) \Pi^c_J\e^{IJ} + (w \G_c\p_I\l) \Pi^c_J\e^{IJ} $$
$$-i\e^{IJ}(w\G_c\p_I\t)(\l\G^c\p_J\t) +ie^{IJ}(w_\a\p_I\t^\a)
(\l_\b\p_J\t^\b)$$
$$+ e^I (\widetilde P_c\Pi_I^c +d_\a\p_I\t^\a +w_\a\p_I\l^\a)]$$
where $\widetilde P_c=P_c +\half B^{flat}_{cAB}\p_I Z^A\p_J Z^B\e^{IJ}$,
$d_\a$ is defined as in \diracsm, and $\widetilde P_c$ and
$d_\a$ are defined to be invariant under supersymmetry
transformations. 
One can easily check that this action reduces to the d=11 superparticle
action and Type IIA superstring action in the appropriate limits. Although
the third line of \spuresma\
vanishes in these limits, the presence of the third
line
will be necessary for BRST invariance of the action.
Note that the first and fourth lines of \spuresma\ simplify to
$P_c\widehat\p_0 x^c + p_\a \widehat\p_0\t^\a + w_\a \widehat\p_0\l^\a$
where $\widehat\p_0=\p_0+e^I\p_I$
when written in terms of the non-supersymmetric variables $P_c$ and $p_\a$.
However, unlike the superstring action, the second line of \spuresma\
which comes
from the supermembrane Hamiltonian
remains complicated when written in terms of $P_c$ and $p_\a$.

As in the d=11 superparticle, the supermembrane action of \spuresma\ needs
to be supplemented with the BRST-like constraint $Q=\l^\a d_\a$.
Using the canonical commutation relations of \antsmtwo, this constraint
generates the BRST transformation
\eqn\brstf{Q\t^\a=\l^\a,\quad Q x^c= {i\over 2}\l\G^c\t,\quad
Qd_\a = -i\widehat \Pi_0^c(\G_c\l)_\a +{i\over 2}\e^{IJ}\Pi_{Ic}
\Pi_{Jd}(\G^{cd}\l)_\a,\quad Qw_\a= d_\a,}
where the equation of motion $\widetilde P^c=\widehat\Pi_0^c\equiv
\Pi_0^c + e^I\Pi_I^c$ has been used.
In addition, in order to allow the construction of a BRST-invariant
action, it will be assumed that the Lagrange multipliers transform under
a BRST transformation as
\eqn\ltra{Q e^I= -i\e^{IJ}\l_\a \p_J\t^\a.} The necessity of \ltra\
can be seen from the kappa transformation of \smak, and differs from the
superstring kappa and BRST transformations of \iiak\ and \flatiit\ where
the Lagrange multipliers are invariant. This difference comes from the
fact that supermembrane kappa transformations do not preserve the Type IIA
superstring condition that $x^{11}=\tau_2$. So to restore the condition
$x^{11}=\tau_2$ after performing a kappa or BRST transformation, one
needs to perform a worldvolume reparameterization of $\tau_2$ which
transforms the Lagrange multipliers $e^I$.

A second important difference between the supermembrane 
BRST transformations
of \brstf\ is that $\l\G^c\l=0$ is not enough to guarantee that
$Q$ is nilpotent. Although $Q^2\t^\a=Q^2 x^c=0$ when $\l\G^c\l=0$,
\eqn\qsqd{Q^2 d_\a = (\l\G^c\widehat\p_0\t)(\G_c\l)_\a -\e^{IJ}
(\l\G_c\p_I\t) \Pi_{Jd}(\G^{cd}\l)_\a +\e^{IJ}(\l_\b\p_I\t^\b)
\Pi_J^c (\G_c\l)_\a }
$$ = - (\l\G^c\G^d\p_I\t)\Pi_{Jd}\e^{IJ}(\G_c\l)_\a -\e^{IJ}
(\l\G_c\p_I\t) \Pi_{Jd}(\G^{cd}\l)_\a +\e^{IJ}(\l_\b\p_I\t^\b)
\Pi_J^c (\G_c\l)_\a $$
$$ = -(\l\G^{cd}\p_I\t)\Pi_{Jd}\e^{IJ}(\G_c\l)_\a -\e^{IJ}
(\l\G_c\p_I\t) \Pi_{Jd}(\G^{cd}\l)_\a $$
$$= \half (\l\G_{cd}\l)\Pi_I^c \e^{IJ}(\G^d\p_J\t)_\a,
$$
where the equation of motion $\nabla\t^\a=0$ has been used and
\eqn\nabthe{\nabla\t^\a =(\p_0+e^I\p_I)\t^\a+(\G^c\p_I\t)^\a\Pi_{Jc}\e^{IJ}.}
Furthermore, $Q^2 e^I= -i\e^{IJ}\l_\a \p_J\l^\a$.

For this reason, the pure spinor constraint $\l\G^c\l=0$ needs
to be supplemented with the additional constraints 
$(\l\G^{cd}\l) \Pi_{Jd}=0$ and $\l_\a\p_J\l^\a=0$ in order that
the BRST operator $Q=\l^\a d_\a$ is nilpotent. The fact that additional
constraints are required for the supermembrane is not surprising since
the supermembrane Hamiltonian is not quadratic, implying that $\l^\a$
is not a free worldvolume variable. So the primary constraint
$\l\G^c\l=0$ does not commute with the supermembrane Hamiltonian and
needs to be supplemented with secondary constraints using the standard
Dirac procedure. 
To find these secondary constraints, note that under $\d w_\a=
(\G^c\l)_\a \Lambda_c,$
\eqn\secondf{\d S_{pure}=\int d^2\tau \Lambda_c[
\half \widehat\p_0(\l\G^c\l) +(\l\G^c\G^d\p_I\l)\Pi_{Jd}\e^{IJ}
-i\e^{IJ}(\l\G^{cd}\p_I\t)(\l\G_d\p_J\t)]}
$$ =\int d^2\tau \Lambda_c[
\half \widehat\p_0(\l\G^c\l) +(\l_\a\p_I\l^\a)\Pi_J^c\e^{IJ}
+\half\p_I[(\l\G^{cd}\l)\Pi_{Jd}\e^{IJ}]+ {i\over 4}(\l\G^d\l)
(\p_I\t\G^{cd}\p_J\t)\e^{IJ}].$$
So $(\l\G^{cd}\l)\Pi_{Jd}=0$ and $\l_\a\p_J\l^\a=0$ are an appropriate
choice of secondary constraints.

Note that the secondary constraints $(\l\G^{cd}\l)\Pi_{Jd}=0$
and $\l_\a\p_J\l^\a=0$ also do not commute with the Hamiltonian, and therefore
lead to further secondary constraints. However, one can easily check
that the complete set of primary and secondary constraints generated
in this manner is first-class. Furthermore, for checking nilpotence
of $Q$ and BRST invariance of the supermembrane action, only
\eqn\consma{\l\G^c\l=0,\quad (\l\G^{cd}\l)\Pi_{Jc}=0, \quad \l_\a\p_J\l^\a=0}
are required. 

The most direct way to check BRST invariance of the supermembrane action
is to show that \spuresma\ is invariant under \brstf\ and \ltra\
up to the constraints of \consma.
However, a more elegant way to show BRST invariance is to use
the Oda-Tonin method of \tonin\ and write the action of \spuresma\ as 
\eqn\otsm{S_{pure}=S+\int d^3\tau ~Q[w_\a \nabla\t^\a]}
where $\nabla\t^\a$ is defined in \nabthe\ and
$S$ is the standard supermembrane action of \sma\ in the 
reparameterization gauge $e^0=-\half$.
Using \trsmak\ and replacing $\xi^\a$ with $\l^\a$, one finds that
\eqn\repfind{Q S_{pure}=QS + \int d^3\tau ~Q^2[w_\a\nabla\t^\a]}
$$=i\int d^3\tau
[\l^\a(\G^c_{\a\b}P_c-\half
\G^{cd}_{\a\b}\Pi_{Ic}\Pi_{Jd}\e^{IJ})\nabla\t^\b + Q^2(w_\a) 
\nabla\t^\a] =0,$$
so $S_{pure}$ is BRST invariant. 

\subsec{Coupling the supermembrane to a d=11 supergravity background}

By starting with the standard supermembrane action in a curved background
and using the background-dependent version of the BRST transformations,
one can also use the Oda-Tonin method to construct the BRST-invariant
version of the supermembrane action in a curved background. In a curved
background, the BRST transformations of \brstf\ and \ltra\ are generalized to
\eqn\backbrst{\widehat Q Z^B= E^B_\ulaa \l^\ulaa,\quad
\widehat Q d_\ulaa = -
 i\widehat \Pi_0^\ulc(\G_\ulc\l)_\ulaa +{i\over 2}\e^{IJ}\Pi_{I\ulc}
\Pi_{J\uld}(\G^{\ulc\uld}\l)_\ulaa,\quad \widehat Qw_\ulaa= d_\ulaa,}
$$\widehat Q e^I= -i\e^{IJ}\l_\ulaa E^\ulaa_A\p_J Z^A.$$

For $\widehat Q$ to be nilpotent in a curved background, one needs
to impose the constraints
\eqn\consmac{\l\G^\ulc\l=0,\quad (\l\G^{\ulc\uld}\l)\Pi_{J\ulc}=0, 
\quad \l_\ulaa\nabla_J\l^\ulaa =0}
where 
\eqn\defnablal{\nabla_J\l^\ulaa\equiv\p_J\l^\ulaa + \Omega_B^{\ulc\uld}
(\G_{\ulc\uld}\l)^\ulaa \p_J Z^B
+\l^\ulbb T_{\ulbb \ulc}^\ulaa \Pi^\ulc_J,}
$T_{\ulbb \ulc}^\ulaa$ is
a dimension-one component of the superspace torsion which is related
on-shell to the four-form field strength $H_{\ula\ulb\ulc\uld}$,
and $\Omega_B^{\ulc\uld}$ is the superspace spin connection. Note that
the third constraint of \consmac\ is obtained from requiring that
$\widehat Q^2 e^I=0$.

The resulting BRST-invariant action in a curved background is
\eqn\resultc{\widehat S_{pure}=\widehat S+
\int d^3\tau \widehat Q[w_\ulaa \nabla \Theta^\ulaa]}
where 
\eqn\defne{\nabla \Theta^\ulbb=
E^\ulbb_A \widehat\p_0 Z^A +\G_{\ula\ulgg}^{\ulbb}E^{\ulgg}_A
\p_I Z^A\Pi_J^\ula\e^{IJ}}
and $\widehat S$ is defined in \smacurved\ after gauge-fixing $e^0=-\half$ and
setting $P^\ulc=\widehat\Pi_0^\ulc$. Using the background-dependent
BRST transformation of \backbrst, one finds
\eqn\findspurec{\widehat S_{pure}=\int d^3\tau
[\half\widehat\Pi_0^{\ulc} \widehat\Pi_{0\ulc} + B_{ABC}
\p_{[0}Z^A\p_{1}Z^B \p_{2]} Z^C
-\half \det(\Pi_I^\ulc\Pi_{J\ulc}) 
+ d_\ulaa \nabla \Theta^\ulaa }
$$+w_\ulaa(\widehat\nabla_0\l^\ulaa +\G_{\ulc\ulbb}^\ulaa
(\nabla_I\l^\ulbb)\Pi_J^\ulc \e^{IJ}) $$
$$-i\e^{IJ}(w_\ulaa \G^\ulaa_{\ulc\ulbb} E^\ulbb_B \p_I Z^B)
(\l_\ulgg \G_\uldd^{\ulc\ulgg} E^\uldd_A \p_J Z^A) 
+i\e^{IJ}(w_\ulaa E^\ulaa_B \p_I Z^B) (\l_\ulgg E^\ulgg_A \p_J Z^A)]$$
where $\nabla_0\l^\ulbb$ and
$\nabla_I\l^\ulbb$ are defined as in \defnablal.

At first sight, it might seem surprising that the pure spinor version
of the supermembrane action in a curved background does not reduce
to the Type IIA superstring sigma model action of \iicpure\
after double-dimensional
reduction. Although both actions are BRST invariant, the $\l\G^{11}\l=0$
constraint in the supermembrane formalism implies that the curved
background fields couple differently in the two actions.
For example, there is no analog of the Fradkin-Tseytlin
term $\a'\int d^2\tau\Phi r$ or the $\int d^2\tau R^{\ulmu\ulhatnu} d_{L\ulmu}
d_{R\ulhatnu}$ term in the supermembrane action. However, since the
Type IIA superstring sigma model is only valid for perturbative
string theory,
these two actions are only guaranteed to agree in the limit when
the string coupling constant goes to zero. In fact, it is clear that
the Fradkin-Tseytlin term $\a'\int d^2\tau\Phi r$ is not possible
in the supermembrane action since there are no scalars which can
play the role of the dilaton. Also, the term $\int d^2\tau
R^{\ulmu\ulhatnu} d_{L\ulmu} d_{R\ulhatnu}$ vanishes when the string
coupling constant goes to zero since $R^{\ulmu\ulhatnu}$ is proportional
to $e^\phi F^{\ulmu\ulhatnu}$ where
$F^{\ulmu\ulhatnu}$ is the Ramond-Ramond field strength. As will
be discussed in the following section, in order to relate the supermembrane
with the Type IIA superstring at non-zero string coupling constant,
one has to compute supermembrane
scattering amplitudes.

\newsec{Supermembrane Scattering Amplitudes}

In this section, a prescription is given for computing scattering
amplitudes using the supermembrane action. These amplitudes might
be useful for studying non-perturbative properties of the Type IIA
superstring.

\subsec{Supermembrane massless vertex operators}

In order to compute scattering amplitudes using the supermembrane
action, one first needs to define BRST-invariant integrated and
unintegrated massless supermembrane vertex operators. 
As in the superstring sigma model
action of \iicpure, the massless integrated supermembrane
vertex operator $\int d^3\tau V$ can be defined as the linear
contribution of the background superfields to 
$\widehat S_{pure}$ of \findspurec. So $V$ has the form
\eqn\intmem{V=A_{BC} \widehat\p_0 Z^B\widehat\p_0 Z^C 
+ A_{BCD}\widehat\p_0 Z^B \p_I Z^C \p_J Z^D
\e^{IJ} + A_{BCDE} \p_I Z^B\p_J Z^C \p_K Z^D \p_L Z^E  \e^{IJ}\e^{KL}}
$$+ (C^\a_B\widehat\p_0 Z^B + C^\a_{BC}\p_I Z^B\p_J Z^C \e^{IJ})d_\a$$
$$
+ (\Omega^\a_{\b C}
\widehat\p_0 Z^C + \Omega^\a_{\b CD}\p_I Z^C\p_J Z^D \e^{IJ})
w_\a \l^\b + Y^\a_{\b B}\p_I Z^B w_\a \p_J\l^\b \e^{IJ}$$
where the relations between the various superfields in \intmem\ are
determined by the BRST invariance condition that $\int d^3\tau QV=0$.
Note that since the action
$\widehat S_{pure}$ is invariant under the
background-dependent BRST transformations of \backbrst, $\int d^3\tau V$
is invariant up to equations of motion under the flat BRST transformations
of \brstf\ and \ltra. 

To obtain the massless unintegrated supermembrane vertex operator,
one can use the supermembrane version of the method used in subsection
(5.3) for
the superstring. Since $\int d^3\tau QV=0$, 
$QV= \p_i W^i$ where $W^i$ are ghost number one states for $i=0$ to 2.
And since $Q^2 V=0$, $QW^i = \e^{ijk}\p_j Y_k$ where $Y_k$ are ghost
number two states. Finally, since $Q^2 W^i=0$, $Q Y_k= \p_k U$ where
$U$ is a ghost number three BRST-invariant
state which will be identified with the
unintegrated supermembrane vertex operator.
Using this method for the integrated massless vertex operator of \intmem,
one
finds that $U$ is the ghost number three d=11 superparticle wavefunction
$\Psi(\l,x,\t)=\l^\a\l^\b\l^\g A_{\a\b\g}(x,\t)$ of \elewf. 
So one sees that d=11
supergravity fields carry ghost number three since they couple
to the three-dimensional supermembrane worldvolume, while d=10
super-Maxwell and Type II supergravity fields carry ghost number one
and two since they couple respectively to the one-dimensional 
superparticle worldline and two-dimensional superstring worldsheet.

\subsec{Supermembrane scattering amplitude prescription}

To compute supermembrane scattering amplitudes using these integrated
and unintegrated massless vertex operators, one naively should 
evaluate correlation functions of these vertex operators on a supermembrane
worldvolume. However, since there is no $SL(2,R)$ or $SL(2,C)$ of the
supermembrane worldvolume, it is not obvious how many vertex operators
should be unintegrated and how many should be integrated. Furthermore,
since the supermembrane action is not conformally invariant, it is
not clear what type of worldvolumes should be included in the correlation
function.

One natural requirement is that the worldvolume zero modes in the 
correlation function should be normalized using the d=11 superparticle
norm $\langle (\l^7\t^9)\rangle$, where the spinor index
contractions of $\l^\a$ and $\t^\b$ in $(\l^7\t^9)$ are those of
\elfact. Since the state $(\l^7\t^9)$ 
is in the cohomology of $Q$ and cannot be written as the supersymmetric
variation of any state in the cohomology of $Q$, use of this
zero mode normalization guarantees that the scattering amplitudes
are gauge invariant and supersymmetric when the external states are on-shell.

The fact that $(\l^7\g^9)$ carries ghost number seven implies that
scattering amplitudes must involve more vertex operators than
just the integrated vertex operator $V$ of ghost number zero and
the unintegrated vertex operator $U$ of ghost number three.
It will now be argued that a correct prescription for an $N$-point
supermembrane scattering amplitude is to use $N-2$ 
integrated vertex operators $V$ of ghost number zero, one unintegrated
vertex operator $U$ of ghost number three, and one unintegrated 
vertex operator $U^*$ of ghost number four. For massless external
states, the ghost number three unintegrated vertex operator is
the d=11 superparticle wavefunction $U=\l^\a\l^\b\l^\g A_{\a\b\g}(x,\t)$
for the linearized supergravity fields, while the ghost number four
unintegrated vertex operator is the d=11 superparticle wavefunction
$U^*=\l^\a\l^\b\l^\g\l^\d A^*_{\a\b\g\d}(x,\t)$ for the linearized
supergravity antifields.

Although this amplitude prescription may sound unusual, it will
be shown below that it can also be used for $N$-point open and closed
string tree amplitudes.
The prescription in some sense resembles the old operator formalism
for computing scattering amplitudes where two external strings
are treated as initial and final states, and the remaining $N-2$
external strings are treated as vertices. Using this interpretation,
the open or closed string worldsheet is an infinitely long strip or
cylinder of zero curvature, and the vertices represent infinitesimally
short strings. So when used for supermembrane scattering amplitudes,
this prescription only involves supermembrane worldvolumes of zero
curvature. Note that unlike superstring amplitudes, one does not
expect to expand over worldvolumes of different genus for supermembrane
amplitudes since there is no scalar spacetime field whose expectation
value could play the role of a dimensionless coupling constant.

In open string theory, the
analogous prescription involves $N-2$ integrated vertex operators,
one ghost number one unintegrated vertex operator $U$ for the string
field, and one
ghost number two integrated vertex operator $U^*$ for the string antifield. 
In bosonic open string theory, these vertex operators can be chosen
in the Siegel gauge to satisfy $U=cV$ and $U^*= c\p c V$ where
$V$ is a dimension one
primary field which is independent of the $(b,c)$ ghosts.
Computing the correlation function
\eqn\corrbos{{\cal A}=\langle U_1(z_1) U^*_2(z_2)\int dz_3 V_3(z_3) ...
\int dz_N V_N(z_N)\rangle,}
one obtains 
\eqn\obtainbos{{\cal A}=(z_1-z_2)^2
\langle V_1(z_1) V_2(z_2)\int dz_3 V_3(z_3) ...
\int dz_N V_N(z_N)\rangle.}
Since \obtainbos\ is invariant under the $SL(2,R)$ transformation
$z_r\to (az_r+b)/(c z_r +d)$, one can fix $(z_1,z_2,z_3)$ so that
the integral over $\int dz_3$ gives a trivial constant infinite factor
which is independent of the
external vertex operators. After dividing by this infinite constant
factor, one recovers the standard open string $N$-point tree amplitude
expression
\eqn\recovbos{{\cal A}=(z_1-z_2)(z_3 - z_1)(z_3-z_2)
\langle V_1(z_1) V_2(z_2) V_3(z_3) \int dz_4 V_4(z_4) ...
\int dz_N V_N(z_N)\rangle.}

In closed string theory, the analogous prescription for $N$-point
tree amplitudes involves $N-2$ integrated vertex operators, one
ghost number two unintegrated vertex operator $U$ for the string field, and
one ghost number four unintegrated vertex operator $U^*$ for the string
antifield. In bosonic closed string theory in Siegel gauge, these
unintegrated vertex operators are $U=c_L c_R V$ and $U^* = c_L(\p_L c_L)
c_R(\p_R c_R) V$ where $V$ is a dimension $(1,1)$ primary field which
is independent of the ghosts. As in open string amplitudes, $SL(2,C)$ 
invariance of the amplitude 
\eqn\corrcbos{{\cal A}=\langle U_1(z_1) U^*_2(z_2)\int d^2 z_3 V_3(z_3) ...
\int d^2 z_N V_N(z_N)\rangle}
implies that the $\int d^2 z_3$ integral provides a trivial constant
infinite factor. 
After dividing out this infinite factor, one recovers
the standard expression for the closed string $N$-point tree amplitude.

Using the pure spinor version of open superstring field theory with
massless external states, the unintegrated ghost number one vertex
operator is $U=\l^\mu A_\mu(x,\t)$ and the ghost number two vertex
operator is $U^*=\l^\mu\l^\nu A^*_{\mu\nu}(x,\t)$
where $A_\mu$ depends on the super-Maxwell photon $a_m$ and photino
$\chi^\mu$ satisfying \phot\ and
$A^*_{\mu\nu}$ depends on the super-Maxwell antiphoton $a^*_m$ and antiphotino
$\chi^*_\mu$ satisfying \antieom. 
The analog of Siegel gauge for these fields
and antifields is 
\eqn\siego{a_m = a^*_m, \quad \chi^\mu = \g_m^{\mu\nu}\p^m\chi^*_\nu}
using the gauge-fixing conditions 
\eqn\gfc{\p^m a_m = 0,\quad
\p_n\p^n a^*_m= \p^n\p_n \chi^*_\nu=0.}
Note that the gauge-fixing
conditions on the fields are the antifield equations of motion, while
the gauge-fixing conditions on the antifields is that they are annihilated
by $\p^n\p_n$. Also note that the identification of \siego\ is consistent
with the equations of motion of \phot\ and \antieom\ in this gauge.
So after dividing out the constant factor and using \siego\ to map
antifields into fields, open superstring massless tree amplitudes
can be computed using this prescription.

Using the pure spinor version of closed superstring field theory with
massless external states, the unintegrated ghost number two vertex
operator is $U=\l_L^\mu\l_R^{\hat\nu} A_{\mu\hat\nu}(x,\t_L,\t_R)$ 
and the ghost number four vertex
operator is $U^*=\l_L^\mu\l_L^\nu\l_R^{\hat\rho}\l_R^{\hat\sigma}
A^*_{\mu\nu\hat\rho\hat\sigma}(x,\t_L,\t_R)$
where $A_{\mu\hat\nu}$ depends on the Type II supergravity fields
$[a_{mn},\chi_{Lm}^{\hat\mu},\chi_{Rm}^\mu,F^{\mu\hat\nu}]$ satisfying
\compeom\ and \compga, and 
$A^*_{\mu\nu\hat\rho\hat\sigma}$ depends on their antifields
$[a^*_{mn},\chi^*_{Lm\hat\mu},\chi^*_{Rm\mu},F^*_{\mu\hat\nu}]$.
The analog of Siegel gauge for these Type II supergravity fields and
antifields is
\eqn\siegc{a_{mn} = a^*_{mn}, \quad \chi_{Lm}^{\hat\mu} = 
\g_n^{\hat\mu\hat\nu}\p^n\chi^*_{Lm\hat\nu},\quad
\chi_{Rm}^\mu = 
\g_n^{\mu\nu}\p^n\chi^*_{Rm\nu},\quad
F^{\mu\hat\nu}=
\g_m^{\mu\rho}\g_n^{\hat\nu\hat\sigma}\p^m\p^n F^*_{\rho\hat\sigma},}
using the gauge-fixing conditions 
\eqn\gfcc{\p^m a_{mn}=\p^n a_{mn}=
\p^m\chi_{Lm}^{\hat\mu}=\p^m\chi_{Rm}^\mu = 0,}
$$\p_n\p^n a^*_{mn}= \p^n\p_n \chi^*_{Lm\hat\nu}=
\p^n\p_n \chi^*_{Rm\nu}=\p^n\p_n F^*_{\mu\hat\nu}=0.$$
The relations of \siegc\ and \gfcc\ can be understood as the ``left-right''
product of the relations of \siego\ and \gfc.
So after dividing out the constant factor and using \siegc\ to map
antifields into fields, open superstring massless tree amplitudes
can be computed using this prescription.

Finally, using the pure spinor version of supermembrane theory with
massless external states,
the unintegrated ghost number three vertex
operator is $U=\l^\a\l^\b\l^\g A_{\a\b\g}(x,\t)$ 
and the ghost number four vertex
operator is $U^*=\l^\a\l^\b\l^\g\l^\d A^*_{\a\b\g\d}(x,\t)$ 
where $A_{\a\b\g}$ depends on the d=11 supergravity fields
$[g_{bc},b_{bcd}, \chi_b^\a]$ satisfying \eqel, and
$A^*_{\a\b\g\d}$ depends on their antifields
$[g^*_{bc},b^*_{bcd}, \chi^*_{b\a}]$ satisfying \elantif.
The analog of Siegel gauge for these d=11 supergravity fields and
antifields is
\eqn\siegm{g_{bc} = g^*_{bc}, \quad  b_{bcd}=b^*_{bcd},
\quad \chi_b^\a= \G_c^{\a\b}\p^c \chi^*_{b\b} +{1\over 9}(\G_b\G^c\G^d)^{\a\b}
\p_c \chi^*_{d\b}}
using the gauge-fixing conditions 
\eqn\gfcc{\p^b g_{bc}-\half\p_c(\eta^{de}g_{de})= \p^b b_{bcd}=
\p^b\chi_b^\a=0,\quad 
\p^b\p_b g^*_{cd}=\p^b\p_b b^*_{cde}=\p^b\p_b\chi^*_{c\a}=0.}
The identification for the gravitino and its antifield in \siegm\ can
be obtained by requiring consistency of the gravitino
equation of motion with the gauge-fixing
conditions of \gfcc. 
So as in the open and closed superstring, it should be possible to
use the map of \siegm\ to obtain d=11 supergravity amplitudes from
the supermembrane amplitude prescription.

\subsec{M-theory conjecture for supermembrane amplitudes}

Having shown in the previous subsection that the prescription for
supermembrane scattering amplitudes has an analog for the superstring,
it is natural to ask if there is some relation between
the supermembrane and superstring scattering amplitudes. It will now
be conjectured that after compactification of $x^{11}$
on a circle of radius $R_{11}$, the supermembrane massless scattering
amplitude coincides with the non-perturbative Type IIA superstring
massless scattering amplitude with the string coupling constant equal
to $(R_{11})^{3\over 2}$. This conjecture is based of course on the
M-theory conjecture of \wittm.

Since the supermembrane action of \spuresma\ is not quadratic, it will not
be possible to obtain exact expressions for correlation functions
of supermembrane vertex operators as was done for correlation functions
of superstring vertex operators. However, since the supermembrane action
reduces in the infinite tension limit to the d=11 superparticle action
of \pureactionel\
which is quadratic, it might be possible to compute supermembrane
scattering amplitudes as a perturbative expansion in the inverse of the
membrane tension. Hopefully, the non-renormalizability of the supermembrane
action will not be an insurmountable obstacle in performing this
perturbative expansion. Since the expansion in the membrane tension is
manifestly d=11 super-Poincar\'e covariant, even the lowest order terms
in the expansion will contain information about Type IIA superstring
amplitudes that is non-perturbative in the string coupling constant.

The first step in studying this M-theory conjecture is to get a better
understanding of the relation beween the supermembrane massless vertex
operators and the Type IIA superstring massless vertex operators. Although
the supermembrane action in a flat background reduces to the
Type IIA superstring action in a flat background after double-dimensional
reduction of $x^{11}$, the 
supermembrane action in a curved background does not reduce to the
Type IIA superstring action in a curved background. Since the integrated
version of the massless vertex operator comes from the linearized
contribution of the curved background, this means
that the integrated version of the supermembrane
massless vertex operator does not reduce to the integrated version
of the Type IIA massless vertex operator. For example, there is no
term quadratic in $d_\a$ in the massless
integrated supermembrane vertex operator
of \intmem\ which reduces to the $d_{L\mu} d_{R\hat\nu} R^{\mu\hat\nu}$
term in the massless integrated Type IIA superstring vertex operator. 
As discussed earlier, this difference comes from the $\l\G^{11}\l=0$
constraint in the supermembrane formalism which is not present in
the Type IIA superstring formalism.

Also, the unintegrated supermembrane vertex operators $U_{membrane}$
and $U^*_{membrane}$ of ghost number three and four are different from the
unintegrated Type IIA superstring vertex operators $U_{string}$
and $U^*_{string}$ of ghost number two and four. However, in this case,
there is a simple way to relate the supermembrane and superstring
vertex operators as will now be shown.

Although $Q U_{membrane}=0$ when $\l\G^c\l=0$ for $c=1$ to 11,
it is not necessarily zero if $\l\G^m\l=0$ for $m=1$ to 10 but
$\l\G^{11}\l$ is non-zero. In this case,
$QU_{membrane}=(\l\G^{11}\l)Y$ where $Y$ is some ghost number two
state annihilated by $Q$. Note that because of \argue, $Y$ is defined
up to terms proportional to $\l\G^{m~11}\l=\l_L\g^m\l_L-\l_R\g^m\l_R$.
Since $\l_L\g^m\l_L-\l_R\g^m\l_R=0$ using the pure spinor version
of the Type IIA superstring, one can identify $Y$ with $U_{string}$.
In other words,
after double-dimensional reduction,  
\eqn\memstr{QU_{membrane}=(\l\G^{11}\l)U_{string}.}

To relate $U^*_{string}$ and $U^*_{membrane}$, first consider
$(Q_L+Q_R)U^*_{string}$ when 
$\l_L\g^m\l_L+\l_R\g^m\l_R=0$ 
but $\l_L\g^m\l_L-\l_R\g^m\l_R$ is non-zero. In this case, 
\eqn\memstar{(Q_L+Q_R)U^*_{string}=
(\l_L\g^m\l_L-\l_R\g^m\l_R) S_m}
where $S_m$ is
some ghost number three vector state which, because of \argue, is defined
up to terms proportional to $\l\G^{11}\l=\l_L^\mu \l_{R\mu}$. 
Furthermore, $(Q_L+Q_R)^2 U^*_{string}=0$ and 
$\l_L\g^m\l_L+\l_R\g^m\l_R=0$ implies that 
\eqn\memstar{(Q_L+Q_R) S_m= (\l_L\g_m)_\mu T_L^\mu +
(\l_R\g_m)^\mu T_{R\mu}}
where $T_L^\mu$ and $T_{R\mu}$ are some ghost number three spinor states
which are defined up to
\eqn\uptos{\d T_L^\mu  = a\l_L^\mu+ b_m (\g^m\l_R)^\mu + 
c^{np}(\g_{np}\l_L)^\mu,\quad
\d T_{R\mu}  = a\l_{R\mu}+ b_m (\g^m\l_L)_\mu + 
c^{np}(\g_{np}\l_R)_\mu.}
After double dimensional reduction, 
\eqn\starmem{U^*_{membrane}=\l_L^\mu T_{R\mu} +\l_{R\mu} T_L^\mu,}
which is invariant under \uptos\ up to terms proportional to
$\l\G^c\l$ for $c=1$ to 11. Furthermore, $(Q_L+Q_R)^2 S_m=0$
implies that $Q U^*_{membrane}=0$ up to terms proportional to
$\l\G^c\l$. 

So, in this way, the unintegrated supermembrane and superstring
vertex operators can be related to each other. For studying
the M-theory conjecture for supermembrane scattering amplitudes,
it would be useful to find a similar relation between the 
integrated supermembrane and superstring vertex operators.

\newsec{Appendix: Zero Momentum Cohomology of d=11 Superparticle}

In this appendix, the zero momentum cohomology of the d=11
superparticle BRST operator, $Q=\l^\a d_\a$, will be 
computed for arbitrary ghost number and shown to correspond
to the linearized
d=11 supergravity ghosts, fields, antifields and antighosts.
As in the case of the N=1 d=10 superparticle discussed in the appendix
of \puresuperp, the zero momentum cohomology of $Q$ is equivalent to the
``linear'' cohomology of a nilpotent operator $\widetilde Q$ involving
an unconstrained bosonic spinor $\lt^\a$ where
\eqn\wideqel{\widetilde Q = \lt^\a p_\a +
\lt\G^c\lt b_{(-1)c}+ c_{(1)}^c(\lt\G_{cd}\lt u_{(-2)}^d +
\lt\G^d\lt u_{(-2)[cd]}) + ...  .}
The new ghost variables
$[b_{(-n)},c_{(n)}]$ and 
$[u_{(-n)},v_{(n)}]$ are fermionic and bosonic pairs of conjugate
variables carrying ghost number $[-n,n]$ which substitute the pure
spinor constraint on the $\l^\a$ variable, and
``linear'' cohomology signifies states in the cohomology of 
$\widetilde Q$ which are at most linearly dependent on these new
variables. 

Although $\widetilde Q$ for the d=11 superparticle involves more terms
than for the N=1 d=10 superparticle, the proof of equivalence of
its ``linear'' cohomology with the zero momentum cohomology of
$Q=\l^\a d_\a$ is identical to the proof in the appendix of \puresuperp\ and
will not be repeated here. As in the N=1 d=10 superparticle case
described in \defwq, 
the term $\lt\G^c\lt b_{(-1)c}$ in \wideqel\ imposes the pure spinor
constraint and the remaining terms in \wideqel\ 
remove the extra gauge invariances implied
by this constraint.
As will now be shown for the d=11 superparticle, these remaining terms
involve new ghost variables with ghost numbers up to $[-7,7]$.

The complete expression for $\widetilde Q$ for the d=11 superparticle
is 
\eqn\complete{\widetilde Q = \lt^\a p_\a +
\lt\G^c\lt b_{(-1)c}+ c_{(1)}^c(\lt\G_{cd}\lt u_{(-2)}^d +
\lt\G^d\lt u_{(-2)[cd]})  }
$$+v_{(2)}^c((\lt\G_c)_\a b^\a_{(-2)} +\lt\G^d\lt b_{(-3)(cd)})$$
$$
+v_{(2)}^{[cd]}(\half (\lt\G_{cd})_\a b^\a_{(-2)} 
+\eta_{de}\lt\G^{ef}\lt b_{(-3)(cf)}
+\lt\G^e\lt b_{(-3)[cde]})$$
$$+c_{(2)}^\a (-\lt\G^c\lt u_{(-3)c\a}+\half(\lt\G^{cd})_\a (\lt\G_c)^\b 
u_{(-3)d\b}))$$
$$
+\half c_{(3)}^{(de)}(\lt\G_d)^\a u_{(-3)e\a}+
{1\over 4} c_{(3)}^{[def]}(\lt\G_{ef})^\a u_{(-3)d\a}$$
$$+ v_{(3)}^{c\a } b_{(-4)}^{d\b } M_{c\a ~d\b~\g\d}\lt^\g \lt^\d$$
$$+{1\over 4}
u_{(-4)}^{[def]}(\lt\G_{ef})^\a c_{(4)d\a} +
\half u_{(-4)}^{(de)}(\lt\G_d)^\a c_{(4)e\a}$$
$$
+u_{(-5)}^\a (-\lt\G^c\lt c_{(4)c\a}+\half(\lt\G^{cd})_\a (\lt\G_c)^\b 
c_{(4)d\b})$$
$$+b_{(-5)}^{[cd]}(\half (\lt\G_{cd})_\a v^\a_{(5)} +
\eta_{de}\lt\G^{ef}\lt v_{(4)(cf)}
+\lt\G^e\lt v_{(4)[cde]})$$
$$
+b_{(-5)}^c((\lt\G_c)_\a v^\a_{(5)} +\lt\G^d\lt v_{(4)(cd)})$$
$$+ u_{(-6)}^c(\lt\G_{cd}\lt c_{(5)}^d +
\lt\G^d\lt c_{(5)[cd]}) + b_{(-7)}\lt\G^c\lt v_{(6)c},  $$
where 
$ M_{c\a ~d\b~\g\d}$ are fixed coefficients which will be defined below.
Before explaining the origin of the various terms in \complete, it
will be first be checked that the linear cohomology of $\widetilde Q$
corresponds to the zero momentum d=11 supergravity ghosts, fields,
antifields and antighosts.

Since $\lt^\a$ is unconstrained, the term $\lt^\a p_\a$ in \complete\
implies using the standard quartet mechanism that states in the cohomology
of $\widetilde Q$ are independent of $x^c$ and $\t^\a$. So states in
the ``linear'' cohomology are represented by the elements 
\eqn\onetoa{[1,c_{(1)}^c, v_{(2)}^c, v_{(2)}^{[cd]}, c_{(2)}^\a, c_{(3)}^{(cd)},
c_{(3)}^{[cde]}, v_{(3)}^{c\a}, 
c_{(4)}^{c\a}, v_{(4)}^{[cde]},
v_{(4)}^{(cd)}, v_{(5)}^\a, c_{(5)}^{[cd]}, c^c_{(5)}, v_{(6)}^c, c_{(7)}],}
which were shown in subsection (4.3) to correspond to the d=11 
supergravity ghosts, fields, antifields and antighosts. To map these
elements to states in the zero momentum cohomology of $Q$, one needs
to find BRST-closed expressions involving these elements which commute
with $\widetilde Q$. For example, the BRST-closed expression corresponding
to $c_{(1)}^c$ is $c_{(1)}^c-\lt\G^c\t$,
the BRST-closed expression corresponding
to $v_{(2)}^{[bc]}$ is $v_{(2)}^{[bc]}- c_{(1)}^{[b}\lt\G^{c]}\t
+\half(\lt\G^b\t)(\lt\G^c\t)$, and the
BRST-closed expression corresponding
to $v_{(2)}^{c}$ is $v_{(2)}^{c}- c_{(1)d}\lt\G^{cd}\t
-\half(\lt\G^{cd}\t)(\lt\G_d\t)$. The corresponding states in
the zero momentum cohomology of $Q$ are obtained by setting all new
variables to zero and replacing $\lt^\a$ with $\l^\a$, i.e. the
state corresponding to $c_{(1)}^c$ is $-\l\G^c\t$, 
the state corresponding to $v_{(2)}^{[bc]}$ is 
$\half(\l\G^b\t)(\l\G^c\t)$, and the
state corresponding to $v_{(2)}^{c}$ is 
$-\half(\l\G^{cd}\t)(\l\G_d\t)$. 

Returning now to the explanation of the terms in $\widetilde Q$ of \complete,
the second term $\lt\G^c\lt b_{(-1)c}$ enforces the pure spinor constraint
and is invariant under 
$\d b_{(-1)c}=\lt\G^d\lt f_{[cd]} + \lt\G_{cd}\lt g^d$ for arbitrary
gauge parameters $f_{[cd]}$ and $g^d$. The third term in \complete\ fixes
these gauge invariances, but introduces new gauge invariances of
$u_{(-2)}^d$ and $u_{(-2)}^{[cd]}$ which are gauge-fixed by the fourth
and fifth terms in \complete. Note that only the symmetric part of
$b_{(-3)(cd)}$ is needed in the fourth and fifth terms of \complete\
since the antisymmetric part $b_{(-3)[cd]}$ can be absorbed by a
redefinition of $b_{(-2)}^\a \to b_{(-2)}^\a + (\lt\G^{cd})^\a b_{(-3)[cd]}.$

At this point, the structure of $\widetilde Q$ is quite complicated, but
one can use the correspondence between the new ghost variables and the
d=11 supergravity fields to help in the construction of the remaining
terms in $\widetilde Q$. In fact, it will now be conjectured that all terms
in $\widetilde Q$ can be deduced from the known linearized supersymmetry
transformations of the d=11 supergravity ghosts, fields, antifields and
antighosts. Although this is not surprising since the BRST transformations
generated by $\widetilde Q$ must be supersymmetric, a proof has not
yet been constructed for the following
conjecture. Nevertheless, it is straightforward
to check that the conjecture is consistent with all terms that have
been computed in $\widetilde Q$ using the gauge-fixing procedure.

Note that all terms in $\widetilde Q$ are either linear or
quadratic in $\lt^\a$. The first conjecture is
that terms linear in $\lt^\a$ describe the zero momentum linearized
supersymmetry transformations of the d=11 supergravity fields where
$\lt^\a$ plays the role of the supersymmetry parameter. For example,
the terms
$\half c_{(3)}^{(de)}(\lt\G_d)^\a u_{(-3)e\a}$ and ${1\over 4}
c_{(3)}^{[def]}(\lt\G_{ef})^\a u_{(-3)d\a}$ describe the zero momentum
linearized supersymmetry transformations of the d=11 supergravity fields
\eqn\linea{\d^\a g_{(de)}= \half\G_{(d}^{\a\b} \chi_{e)\b} ,\quad
\d^\a b_{[def]}={1\over 4} (\G_{[ef})^{\a\b} \chi_{d]\b}, }
where $c_{(3)}^{(de)}$ is identified with the graviton $g^{(de)}$,
$c_{(3)}^{[def]}$ is identified with the three-form $b^{[def]}$,
and $u_{(-3)d}^\a$ is identified with the gravitino $\chi_d^\a$.

The second conjecture is that the terms in $\widetilde Q$ which are
quadratic in $\lt^\a$ can be deduced from the anticommutator of two 
linearized supersymmetry transformations where $\lt^\a\lt^\b$
plays the role of the supersymmetry parameters in the anticommutator.
If the d=11 supersymmetry algebra were closed off-shell, the anticommutator
of two supersymmetry transformations acting on any supergravity field
would be proportional to a translation, i.e. 
$\{\d_\a,\d_\b\}\phi_I= \G^c_{\a\b} \p_c \phi_I$ for any $\phi_I$.
However, since
the supersymmetry algebra is only closed on-shell, the anticommutator
of two supersymmetry transformations acting on a supergravity field
can contain a term proportional to equations of motion, i.e.
$\{\d_\a,\d_\b\}\phi_I= \G^c_{\a\b} \p_c \phi_I + M_{IJ~\a\b} 
{{\p {\cal S}}\over
{\p\phi_J}}.$ 

For d=11 supergravity fields, $M_{IJ ~\g\d}$ is non-vanishing when
$I$ and $J$ correspond to gravitino fields, i.e.
\eqn\gravvar{
\{\d_\g,\d_\d\}\chi_{c\a}= \G^d_{\g\d} \p_d \chi_{c\a}
 + M_{c\a~d\b ~\g\d} {{\p {\cal S}}\over {\p\chi_{d\b}}} }
where the coefficients 
$ M_{c\a~d\b ~\g\d}$ can be explicitly computed using the 
linearized supersymmetry transformations of the standard d=11
supergravity action. From the second conjecture, this implies the term
$$ v_{(3)}^{c\a } b_{(-4)}^{d\b } M_{c\a ~d\b~\g\d}\lt^\g \lt^\d$$
in \complete\ where $v_{(3)}^{c\a}$ corresponds to the gravitino
$\chi^{c\a}$ and $b_{(-4)}^{d\b}$ corresponds to the 
gravitino antifield $\chi^*_{d\b}$
whose BRST transformation is the linearized
equation of motion ${{\p {\cal S}}\over{\p\chi^{d\b}}}$.

To give another example of the second conjecture, the term
\eqn\thist{
c_{(2)}^\a (-\lt\G^c\lt u_{(-3)c\a}+\half(\lt\G_{cd})_\a (\lt\G^c)^\d 
u^d_{(-3)\d})}
in \complete\ can be deduced from the anticommutator of two supersymmetry
transformations acting on the supersymmetry ghost $\xi_\a$.
Using $\d^\a\rho_c=(\G_c\xi)^\a$ and $\d^\b\xi_\a=
-\half \p_b\rho_c(\G^{bc})^\b_\a$ where $\rho_c$ is
the reparameterization ghost, one finds that 
\eqn\sughost{\{\d_\b,\d_\g\}\xi_\a = \G^c_{\b\g}\p_c\xi_\a
+
(-\G^c_{\b\g}\p_c\xi_\a +\half(\G_{cd})_{\a(\b} \G^{c\d}_{\g)} \p^d\xi_\d).}
So the term
\thist\ in $\widetilde Q$ can be deduced from
\sughost\ where $c_{(2)\a}$ corresponds to the supersymmetry ghost
$\xi_\a$ and $u^d_{(-3)\d}$ corresponds to the 
gravitino $\chi^d_\d$ whose BRST variation is $\p^d\xi_\d$. 

So one can use these two conjectures to deduce all terms in $\widetilde Q$
of \complete, and one can explicitly check that this construction is
consistent with the required gauge-fixing properties of the term in
$\widetilde Q$. Furthermore, one can check that these conjectures are
also consistent with the BRST operator of \defwq\ for d=10
super-Maxwell theory. Note that the terms in the second half of
\complete\ are related to terms in the first half of \complete\ by
exchanging fields with antifields and ghosts with antighosts, i.e.
by exchanging $[b_{(-n)},c_{(n)}]$ with $[v_{(7-n)},u_{(n-7)}]$.
The term 
$ v_{(3)}^{c\a } b_{(-4)}^{d\b } M_{c\a ~d\b~\g\d}\lt^\g \lt^\d$
is invariant under this exchange since 
$M_{c\a ~d\b~\g\d}
=M_{d\b ~c\a~\g\d}$ in order that 
$\{\d_\g,\d_\d\}{\cal S}= {{\p {\cal S}}\over{\p\chi^{c\a}}}
M_{c\a ~d\b~\g\d}
{{\p {\cal S}}\over{\p\chi^{d\b}}} =0$ where ${\cal S}$ is the linearized d=11
supergravity action.

\vskip 20pt

{\bf Acknowledgements:}
I would like to thank Paul Howe for
useful discussions
and
CNPq grant 300256/94-9, Pronex grant 66.2002/1998-9, 
and FAPESP grant 99/12763-0 for partial financial support.
This research was partially conducted during the period the author
was employed by the Clay Mathematics Institute as a CMI Prize Fellow.

\listrefs

\end